 \documentclass{emulateapj}

\usepackage{color}

\shorttitle{The physical scale and morphology of ultrared DSFGs}
\shortauthors{Oteo et al.}

\usepackage{natbib}

\begin{document}


\title{Witnessing the birth of the red sequence: the physical scale and morphology of dust emission in hyper-luminous starbursts in the early Universe}


\author{I.~Oteo,\altaffilmark{1,2}
R.\,J.~Ivison,\altaffilmark{2,1}
M.~Negrello,\altaffilmark{3}
I.~Smail,\altaffilmark{4,5}
I.~P\'erez-Fournon,\altaffilmark{6,7}
M.~Bremer,\altaffilmark{8}
G.~De~Zotti,\altaffilmark{9}
S.\,A.~Eales,\altaffilmark{3}
D.~Farrah,\altaffilmark{10}
P.~Temi,\altaffilmark{11}
D.\,L.~Clements,\altaffilmark{12}
A.~Cooray,\altaffilmark{13}
H.~Dannerbauer,\altaffilmark{6,7}
S.~Duivenvoorden,\altaffilmark{14}
L.~Dunne,\altaffilmark{1,3}
E.~Ibar,\altaffilmark{15}
A.\,J.\,R.~Lewis,\altaffilmark{1}
R.~Marques-Chaves, \altaffilmark{6,7}
P.~Mart\'inez--Navajas,\altaffilmark{6,7}
M.\,J.~Micha{\l}owski,\altaffilmark{16}
A.~Omont,\altaffilmark{17,18}
S.~Oliver,\altaffilmark{14}
D.~Riechers,\altaffilmark{19}
D.~Scott\altaffilmark{20} and
P.~van~der~Werf\altaffilmark{21}
}

\affil{$^1$Institute for Astronomy, University of Edinburgh, Royal Observatory, Blackford Hill, Edinburgh EH9 3HJ, U.K.}
\affil{$^2$European Southern Observatory, Karl-Schwarzschild-Str. 2, 85748 Garching, Germany}
\affil{$^3$School of Physics \& Astronomy, Cardiff University, Queen's Buildings, The Parade, Cardiff CF24 3AA, U.K.}
\affil{$^4$Centre for Extragalactic Astronomy, Department of Physics, Durham University, South Road, Durham DH1 3LE, U.K.}
\affil{$^5$Institute for Computational Cosmology, Department of Physics, Durham University, South Road, Durham DH1 3LE, U.K.}
\affil{$^6$IAC, E-38200 La Laguna, Tenerife, Spain}
\affil{$^7$Departamento de Astrofisica, Universidad de La Laguna, E-38205 La Laguna, Tenerife, Spain}
\affil{$^8$H.H. Wills Physics Laboratory, University of Bristol, Tyndall Avenue, Bristol BS8 1TL, U.K.}
\affil{$^9$INAF, Osservatorio Astronomico di Padova, Vicolo Osservatorio 5, I-35122 Padova, Italy}
\affil{$^{10}$Department of Physics, Virginia Tech, Blacksburg, VA 24061}
\affil{$^{11}$Astrophysics Branch, NASA/Ames Research Center, MS 245-6, Moffett Field, CA 94035}
\affil{$^{12}$Physics Department, Blackett Lab, Imperial College, Prince Consort Road, London SW7 2AZ, U.K.}
\affil{$^{13}$Department of Physics and Astronomy, University of California, Irvine, CA 92697}
\affil{$^{14}$Astronomy Centre, Department of Physics and Astronomy, University of Sussex, Brighton BN1 9QH, U.K.}
\affil{$^{15}$ Instituto de F\'{i}sica y Astronom\'{i}a, Universidad de Valpara\'{i}so, Avda. Gran Breta\~{n}a 1111, Valpara\'{i}so, Chile}
\affil{$^{16}$Astronomical Observatory Institute, Faculty of Physics, Adam Mickiewicz University, ul.~S{\l}oneczna 36, 60-286 Pozna{\'n}, Poland}
\affil{$^{17}$UPMC Univ Paris 06, UMR 7095, IAP, 75014, Paris, France}
\affil{$^{18}$CNRS, UMR7095, IAP, F-75014, Paris, France}
\affil{$^{19}$Cornell University, Space Sciences Building, Ithaca, NY 14853}
\affil{$^{20}$Department of Physics and Astronomy, University of British Columbia, 6224 Agricultural Road, Vancouver, BC V6T 1Z1, Canada}
\affil{$^{21}$Leiden Observatory, Leiden University, P.O.~Box 9513, NL-2300 RA Leiden, The Netherlands}

\email{ivanoteogomez@gmail.com}


  
\begin{abstract}

We present high-spatial-resolution ($\sim 0.12''$ or $\approx 800 \, {\rm pc}$ at $z = 4.5$) ALMA 870\,$\mu$m dust continuum observations of a sample of 44 ultrared dusty star-forming galaxies (DSFGs) selected from the {\it H}-ATLAS and {\it Her}MES far-infrared surveys because of their red colors from 250 to 500\,$\mu$m: $S_{500} / S_{250} > 1.5$ and $S_{500} / S_{350} > 1.0$. With photometric redshifts in the range $z \sim 4$--6, our sample includes the most luminous starbursting systems in the early Universe known so far, with total obscured star-formation rates (SFRs) of up to $\sim 4,500 \, M_\odot \, {\rm yr}^{-1}$, as well as a population of lensed, less intrinsically luminous sources.  The lower limit on the number of ultrared DSFGs at 870\,$\mu$m (with flux densities measured from the ALMA maps and thus not affected by source confusion) derived in this work is in reasonable agreement with models of galaxy evolution, whereas there have been reports of conflicts at 500\,$\mu$m (where flux densities are derived from SPIRE). Ultrared DSFGs have a variety of morphologies (from relatively extended disks with smooth radial profiles, to compact sources, both isolated and interacting) and an average size, $\theta_{\rm FWHM}$, of $1.46 \pm 0.41\, {\rm kpc}$, considerably smaller than the values reported in previous work for less-luminous DSFGs at lower redshifts. The size and the estimated gas-depletion times of our sources are compatible with their being the progenitors of the most massive, compact, red-and-dead galaxies at $z \sim 2$--3, and ultimately of local ultra-massive elliptical galaxies or massive galaxy clusters. We are witnessing the birth of the high-mass tail of the red sequence of galaxies.
\end{abstract}

\keywords{galaxy evolution; sub-mm galaxies; dust emission; number counts}

\section{Introduction}
\label{intro}

It has been well known for over fifty years \citep[e.g.][]{Rood1969ApJ...158..657R} that galaxies with quiescent stellar populations in the local Universe are located in a region of the color--magnitude or color--mass diagram known as the `red sequence' \citep{Bower1992MNRAS.254..589B,Bower1992MNRAS.254..601B,Strateva2001,Blanton2003ApJ...594..186B,Baldry2004ApJ...600..681B}.  Later work suggested that the red sequence was already in place at $z \gtrsim 2$ \citep{Labbe2005ApJ...624L..81L,Williams2009ApJ...691.1879W,Brammer2009ApJ...706L.173B}, possibly as early as $z \sim 3$ \citep{Whitaker2011ApJ...735...86W,Nayyeri2014ApJ...794...68N}. The analysis of the stellar populations in so-called `red-and-dead' galaxies at $z \sim 2$--3 suggests that their formation could have taken place at even earlier times, at $z \sim 4$--6, in the form of extreme bursts of star formation \citep{Collins2009Natur.458..603C,Thomas2010MNRAS.404.1775T,Gobat2011A&A...526A.133G,Zeimann2012ApJ...756..115Z,Petty2013AJ....146...77P}.

Until recently, only a handful of dusty star-forming galaxies (DSFGs -- e.g.\ \citealt{Casey2014PhR...541...45C}, also sometimes referred to as submm-selected galaxies, SMGs) had been found at $z \sim 4$ \citep{Capak2008ApJ...681L..53C,Daddi2009ApJ...694.1517D,Coppin2009MNRAS.395.1905C,Knudsen2010ApJ...709..210K,Smolcic2011ApJ...731L..27S,Swinbank2012MNRAS.427.1066S}. This situation changed quickly following the launch of the {\it Herschel Space Observatory} \citep{Pilbratt2010A&A...518L...1P} and its wide-field far-infrared (FIR) surveys, which have enormously increased the number of known DSFGs in the early Universe. This is mainly due to the selection of very high-redshift DSFGs by looking for sources whose flux densities increase from 250 to 500\,$\mu$m, as measured by SPIRE \citep{Griffin2010}: $S_{250} < S_{350} < S_{500}$, where the various selection criteria employed to date have been described by, for example,  \citet{Cox2011ApJ...740...63C,Combes2012A&A...538L...4C,Dowell2014ApJ...780...75D,Ivison2016ApJ...832...78I} and \citet{Asboth2016MNRAS.462.1989A}, but see also \S\ref{section_selection_UR_starbursts}.

Follow-up spectroscopic observations has confirmed the effectiveness of the ultrared criterion to select DSFGs at $z = 4$--6 \citep[e.g.][]{Riechers2013Natur.496..329R,Oteo2016ApJ...827...34O,Asboth2016MNRAS.462.1989A,Riechers2017arXiv170509660R,Fudamoto2017arXiv170708967F,Zavala2017arXiv170709022Z}. In this work, we present ALMA 870\,$\mu$m dust continuum observations at $\sim 0.12''$ resolution (or $\approx 800 \, {\rm pc}$ at $z \sim 4.5$) of a sample of 44 ultrared DSFGs at $z = 4$--6, selected from {\it H}-ATLAS \citep{Eales2010PASP..122..499E} and {\it Her}MES \citep{Oliver2010}, the widest surveys carried out by {\it Herschel}.  The main goals are: 1) to confirm that there exists a significant population of unlensed hyper-luminous DSFGs at $z = 4$--6; 2) to study the number density of such sources, to inform and constrain galaxy formation and evolution models; and 3) to study their sizes and morphologies, to  explore their nature and their likely evolution, hoping to reveal the manner in which the ancestors of today's most massive galaxies and structures were formed. 

This paper is structured as follows: In \S\ref{section_selection_UR_starbursts} we present the selection of ultrared DSFGs at $z \sim 4$--6, the sample studied in this work. \S\ref{section_ALMA_imaging_and_data} presents our ALMA high-spatial-resolution observations. The results of the paper are presented and discussed in \S\ref{section_SFR_most_luminous}, \ref{section_lenses} and \ref{sec_morph_scale_dust}. Finally, the main conclusions are summarized in \S\ref{concluuuuu}. The total infrared luminosities ($L_{\rm IR}$) reported in this work refer to the integrated luminosities between rest-frame 8 and $1000 \,{\rm \mu m}$. The reported SFRs are derived from the $L_{\rm IR}$ assuming a Salpeter initial mass function (IMF) and the classical \cite{Kennicutt1998} calibration, ignoring for now the possibility of a profoundly different IMF in these objects \citep[see][]{Romano2017MNRAS.470..401R}. The sizes and areas are calculated by carrying out 2D elliptical Gaussian fits and using $A = \pi \times R_1 \times R_2$, where $R_1$ and $R_2$ are the major and minor semi-axes of the best-fit elliptical Gaussians, respectively: $R_1 = {\rm FWHM}_{\rm major}/2$ and $R_2 = {\rm FWHM}_{\rm minor}/2$.  We assume a flat Universe with $(\Omega_m, \Omega_\Lambda, h_0)=(0.3, 0.7, 0.7)$. For this cosmology, the sky scale is $\sim 6.6\,{\rm
  kpc}/''$ at $z = 4.5$. 

\section{Ultrared DSFGs at $z > 4$}
\label{section_selection_UR_starbursts}

The dust emission of DSFGs peaks at approximately rest-frame 100--$200 \, {\rm \mu m}$ \citep{Dunne2000MNRAS.315..115D,Farrah2003MNRAS.343..585F,Hwang2010}. As the redshift increases, the dust emission peak shifts to longer wavelengths until at $z \gtrsim 4$ it is located at or beyond 500\,$\mu$m. Thus, looking for galaxies whose flux densities rise from 250 to 350\,$\mu$m and then to 500\,$\mu$m, it is possible to select DSFGs at $z > 4$. These are called `500\,$\mu$m risers', or `ultrared' DSFGs. At the flux density limits accessible to SPIRE, the number density of these sources is relatively low, meaning that wide-area surveys are needed to select them.  The sample of ultrared DSFGs studied in this paper has been taken from \cite{Ivison2016ApJ...832...78I} and \cite{Asboth2016MNRAS.462.1989A}.  We refer the reader to those papers for details on the source selection.  Briefly, our galaxies were selected from the {\it H}-ATLAS\footnote{\url https://www.h-atlas.org} Data Release 1 and 2 \citep[][Smith et al.\ in prep., Maddox et al.\ in prep., Furlanetto et al.\ in prep.]{Valiante2016MNRAS.462.3146V,Bourne2016MNRAS.462.1714B} and the fourth {\it Her}MES data release \footnote{\url http://hedam.oamp.fr/HerMES}, by looking for galaxies with $S_{500} / S_{250} \geq 1.5$ and $S_{500} / S_{350} \geq 1.0$, where $S_{\lambda}$ is the flux density in the SPIRE band centered at $\lambda \,[{\rm \mu m}]$.  The {\it Herschel} maps for the full sample were then inspected visually by several members of the team to exclude blended sources, and checked for contamination by bright synchrotron emission from radio-loud AGN.  With our ultrared criterion, our sample is arguably one of the largest, reddest and most robust available.  To put this beyond doubt, we refined our selection criteria still further via 850\,$\mu$m SCUBA-2 and 870\,$\mu$m LABOCA imaging, which provide better resolution than SPIRE at 500\,$\mu$m, to isolate those whose colors are consistent only with $z > 4$: $S_{870} / S_{500} > 0.4$ \citep[e.g.][]{Ivison2016ApJ...832...78I}. Taken together, these criteria ensure the selection of the most secure, ultrared DSFGs. 


The validity of the ultrared criterion to select DSFGs at $z > 4$ has been proven by our recent ALMA and NOEMA spectral scans in the $2 \, {\rm mm}$ and $3 \, {\rm mm}$ windows. These observations aimed to confirm unambiguously the redshifts of our ultrared DSFGs via detection of multiple CO, [C\,{\sc i}](1--0) and water emission lines.  So far, all spectroscopically confirmed ultrared  DSFGs lie at $z_{\rm spec} > 3.5$ \citep{Riechers2013Natur.496..329R,Asboth2016MNRAS.462.1989A,Riechers2017arXiv170509660R,Fudamoto2017arXiv170708967F}.  Most sources without spectroscopic confirmation have a single emission line close to the center of the $3 \, {\rm mm}$ band, suggesting $z > 3$ if the detected lines correspond to CO transitions (lower-redshift options would be in strong disagreement with the FIR-derived photometric redshifts).  No lines were detected in the remaining sources, likely due to the lack of depth \citep[see, e.g., the discussion in][]{Fudamoto2017arXiv170708967F}. For some of our ultrared DSFGs, $z\sim 6$ is likely, and two sources have indeed been spectroscopically confirmed at $z \sim 6$ \citep{Riechers2013Natur.496..329R,Fudamoto2017arXiv170708967F,Zavala2017arXiv170709022Z}. 

A subsample of 44 equatorial and southern sources from our full sample of ultrared DSFGs at $z \sim 4$--6 selected from {\it H}-ATLAS and {\it Her}MES (see Table~\ref{table_flux_density_and_redshift}) were chosen for ALMA 870\,$\mu$m follow-up observations at high spatial resolution ($\sim 0.12''$ or ${\rm \approx 800 \, pc}$, see \S\ref{section_ALMA_imaging_and_data}). These observations constitute the main dataset that we present and analyze in this paper.  We note that not all the sources in our sample of ultrared DSFGs were observed with ALMA.  Instead, the observations were limited to: 1) sources visible during the required ALMA configuration; 2) sources that could share calibration, to minimize calibration overheads.  This is the reason why, for example, no sources from the GAMA-12 or GAMA-15 {\it H}-ATLAS fields were observed.  Table~\ref{table_flux_density_and_redshift} shows the properties of the sample of ultrared DSFGs studied in this paper.

\begin{table*}
\caption{\label{table_flux_density_and_redshift}Properties of our sample of ultrared DSFGs.}
\centering
\begin{tabular}{ccccccccccccccc}
\hline
IAU name					& Nickname 			& $z$\tablenotemark{(a)} 	&	$S_{250}$	&	$S_{350}$	&	$S_{500}$	&	$S_{870}$	&	$S^{\rm ALMA}_{870}$\tablenotemark{(g)} & ${\rm SFR}$ & $N$\tablenotemark{(j)}  \\
 						&  					& 		&	[mJy	]&	[mJy]		&	[mJy]	&	[mJy]		&	[mJy] & [$M_\odot \, {\rm yr}^{-1}$]  \\
\hline\hline
HerMES\,J003929.5+002424 		&	(RARE) HeLMS\_36		& 						&	$140 \pm 7$	&	$152 \pm 8$	&	$162 \pm 9$	&	$50 \pm 4$	&	$58.5 \pm 0.6$		&	$\sim 7855 / \mu$					& 	$1$		\\
HerMES\,J003814.0$-$002253		&	(RARE) HeLMS\_38		&						&	$73 \pm 7$	&	$119 \pm 8$	&	$123 \pm 9$	&	$56 \pm 4$	&	$48.6 \pm 3.4$		&	$\sim 6526 / \mu$					& 	$1$		\\
HerMES\,J004532.5$-$000124		&	(RARE) HeLMS\_54		&						&	$48 \pm 7$	&	$88 \pm 8$	&	$97 \pm 9$	&	$36 \pm 4$	&	$47.7 \pm 0.9$		&	$\sim 6405$						&	$5$		\\
HerMES\,J002220.8$-$015521		&	HELMS\_RED\_4 		& 	5.161\tablenotemark{(b)} 	& 	$62 \pm 6$ 	& 	$104 \pm 6$ 	& 	$116 \pm 7$ 	& 	$41 \pm 3$	& 	$40.9 \pm 2.8$		& 	$\sim 5774 / \mu$					&	$1$		\\
HerMES\,J000303.9+024113		&	(RARE) HeLMS\_42		&						&	$34 \pm 5$	&	$54 \pm 5$	&	$87 \pm 5$	&	$35 \pm 5$	& 	$42.7 \pm 0.9$ 		&	$\sim 5734 / \mu$ \tablenotemark{(h)}	&	$2$		\\
HATLAS\,J090045.4+004125		&	G09--83808			& 	6.027\tablenotemark{(c)}	& 	$10 \pm 5$ 	& 	$25 \pm 5$ 	& 	$44 \pm 6$ 	& 	$36 \pm 3$ 	& 	$36.5 \pm 2.1$ 		&	$\sim 5635 / \mu$					&	$1$		\\
HATLAS\,J000124.9$-$354212		&	SGP--28124			&	$3.93^{+0.08}_{-0.45}$	& 	$62 \pm 8$ 	& 	$89 \pm 8$ 	& 	$118 \pm 9$ 	& 	$47 \pm 6$ 	&  	$41.5 \pm 1.3$ 		& 	$\sim 5573 / \mu$					&	$1$		\\
HATLAS\,J000607.6$-$322639 	&	SGP--261206			&  	4.242\tablenotemark{(c)} 	& 	$23 \pm 5$ 	& 	$45 \pm 6$ 	& 	$59 \pm 7$ 	& 	$57 \pm 9$ 	& 	$40.1 \pm 1.5$		&	$\sim 5371 / \mu$					&	$1$		\\
HATLAS\,J000624.3$-$323019		&	SGP--93302			& 	$3.91^{+0.27}_{-0.22}$	& 	$31 \pm 4$ 	& 	$61 \pm 5$ 	& 	$62 \pm 6$ 	& 	$37 \pm 4$ 	& 	$39.1 \pm 3.2$ 		& 	$\sim 5250 / \mu$\tablenotemark{(i)}		&	$2$		\\
HATLAS\,J000306.9$-$330248 	&	SGP--196076 			&  	4.425\tablenotemark{(d)} 	& 	$29 \pm 5$ 	& 	$29 \pm 6$ 	& 	$46 \pm 7$ 	& 	$33 \pm 4$ 	& 	$34.6 \pm 2.3$ 		& 	$\sim 4643$						&	$3$		\\
HerMES\,J000727.1+015626		&	HELMS\_RED\_19		& 						&	$53 \pm 5$	&	$72 \pm 5$	&	$82 \pm 5$	&	$34 \pm 5$	& 	$33.3 \pm 2.2$ 		&	$\sim 4471 / \mu$					&	$1$		\\
HerMES\,J003257.1$-$424736 	&	ELAISS1\_7 			&   						& 	$48 \pm 4$ 	& 	$69 \pm 5$ 	& 	$71 \pm 5$	&	$-$			& 	$32.9 \pm 1.2$		& 	$\sim 4418 / \mu$					&	$1$		\\
HerMES\,J000900.6+050709		&	HELMS\_RED\_69  		& 						&	$37 \pm 5$	&	$43 \pm 5$	&	$70 \pm 5$	&	$42 \pm 5$	& 	$30.1 \pm 1.4$		&	$\sim 4042 / \mu$ \tablenotemark{(h)}	&	$1$		\\
HerMES\,J003943.5$-$003955		&	HELMS\_RED\_118		&						&	$32 \pm 6$	&	$57 \pm 6$	&	$74 \pm 7$	&	$26 \pm 4$	&	$29.9 \pm 2.1$		&	$\sim 4015 / \mu$					&	$1$		\\
HerMES\,J043657.5$-$543809 	&	ADFS\_27 			&  	5.655\tablenotemark{(e)}	& 	$15 \pm 5$ 	& 	$19 \pm 6$ 	& 	$24 \pm 3$ 	& 	$25 \pm 2$ 	& 	$26.8 \pm 1.0$		& 	$\sim 3968$						&	$2$		\\
HerMES\,J022656.6$-$032709 	&	XMM\_30 				& 						& 	$30 \pm 5$ 	& 	$50 \pm 8$ 	& 	$55 \pm 7$ 	& 	$28 \pm 2$	& 	$27.8 \pm 2.4$		& 	$\sim 3933 / \mu$					&	$1$		\\ 
HATLAS\,J084937.0+001455		&	G09--81106 			& 	4.53\tablenotemark{(c)} 	& 	$14 \pm 5$ 	& 	$31 \pm 6$ 	& 	$48 \pm 7$ 	& 	$30 \pm 5$ 	& 	$28.4 \pm 0.8$ 		& 	$\sim 3840 / \mu $					&	$1$		\\
HATLAS\,J225432.0$-$323904		&	SGP--317726 			& 	$3.69^{+0.39}_{-0.30}$	& 	$20 \pm 5$ 	& 	$35.1 \pm 5$ 	& 	$40 \pm 6$ 	& 	$19 \pm 3$ 	& 	$26.9 \pm 2.9$ 		& 	$\sim 3612$						&	$1$		\\
HATLAS\,J004223.5$-$334340		&	SGP--354388 			& 	4.002\tablenotemark{(f)}	& 	$17 \pm 6$ 	& 	$40 \pm 7$ 	& 	$54 \pm 8$ 	& 	$64 \pm 11$ 	& 	$24.1 \pm 1.7$		& 	$\sim 3257$						&	$3$		\\
HerMES\,J235808.7+005557		&	HELMS\_RED\_68 		&						&	$55 \pm 5$	&	$74 \pm 6$	&	$76 \pm 7$	&	$27 \pm 7$	& 	$24.1 \pm 3.1$		&	$\sim 3236$						&	$2$		\\
HerMES\,J002737.3$-$020759		&	HELMS\_RED\_31		& 	3.798\tablenotemark{(b)}	& 	$36 \pm 7$	& 	$49 \pm 6$ 	& 	$72 \pm 7$	&	$23 \pm 4$	& 	$22.9 \pm 4.1$		& 	$\sim 3092 / \mu$					&	$1$		\\
HerMES\,J043913.5$-$542546 	&	ADFS\_17 			& 						& 	$23 \pm 6$ 	& 	$48 \pm 7$ 	& 	$52 \pm 8$ 	& 		$-$		& 	$21.1 \pm 0.8$ 		& 	$\sim 2833$						&	$3$		\\
HATLAS\,J084059.3$-$000417 	&	G09--81271			&	$4.62^{+0.46}_{-0.38}$ 	& 	$15 \pm 5$ 	& 	$31 \pm 6$ 	& 	$42 \pm 7$ 	& 	$30 \pm 4$ 	& 	$20.5 \pm 0.7$ 		& 	$\sim 2753 / \mu$					&	$1$		\\
HerMES\,J004724.7+010114		&	HELMS\_RED\_82		&						&	$47 \pm 7$	&	$76 \pm 8$	&	$76 \pm 9$	&	$27 \pm 4$	&	$19.6 \pm 2.3$		&	$\sim 2632 / \mu $					&	$1$		\\
HerMES\,J002851.4$-$431351		&	ELAISS1\_18 			& 	 					& 	$26 \pm 4$ 	& 	$43 \pm 5$ 	& 	$45 \pm 5$ 	& 	$27 \pm 3$ 	& 	$18.2 \pm 0.8$		& 	$\sim 2444 $						&	$2$		\\
HerMES\,J002314.7+001243		&	HeLMS\_182 			& 						& 	$33 \pm 5$ 	& 	$58 \pm 7$ 	& 	$73 \pm 6$ 	& 	$31 \pm 4$ 	& 	$17.7 \pm 0.7$ 		& 	$\sim 2377$						&	$1$		\\
HATLAS\,J000018.0$-$333737		&	SGP--72464			& 	$3.06^{+0.21}_{-0.19}$	& 	$20 \pm 6$ 	& 	$30 \pm 8$ 	& 	$38 \pm 8$ 	& 	$10 \pm 4$ 	& 	$16.9 \pm 0.4$ 		& 	$\sim 2269$						&	$1$		\\
HerMES\,J021914.2$-$043740		&	XMM\_76 				& 						& 	$26 \pm 5$	& 	$37 \pm 8$ 	& 	$38 \pm 6$	&	$17 \pm 2$	& 	$16.8 \pm 1.6$		& 	$\sim 2256 / \mu$					&	$1$		\\
HerMES\,J044509.9$-$530006		&	ADFS\_31 			& 						& 	$19 \pm 7$ 	& 	$34 \pm 7$ 	& 	$40 \pm 7$ 	& 	$25 \pm 5$ 	& 	$16.7 \pm 0.8$		& 	$\sim 2242$						&	$2$		\\
HerMES\,J235830.9$-$005632		&	HELMS\_RED\_270  	& 						&	$48 \pm 5$	&	$62 \pm 5$	&	$64 \pm 5$	&	$29 \pm 5$	& 	$16.5 \pm 0.9$		&	$\sim 2216$						&	$2$		\\
HATLAS\,J010740.9$-$282709		&	SGP--32338 			&	$3.93^{+0.26}_{-0.24}$	& 	$16 \pm 7$ 	& 	$33 \pm 8$ 	& 	$64 \pm 8$ 	& 	$23 \pm 3$	& 	$16.4 \pm 0.9$		& 	$\sim 2202$						&	$2$		\\
HerMES\,J002933.9$-$421212		&	ELAISS1\_40 			&  						&	 $25 \pm 4$ 	& 	$34 \pm 2$ 	& 	$41 \pm 5$	&	$-$			& 	$14.4 \pm 0.6$		& 	$\sim 1934$						&	$1$		\\
HATLAS\,J225855.7$-$312405		&	SGP--499646			&	$4.68^{+0.49}_{-0.34}$	& 	$6 \pm 6$ 	& 	$10 \pm 6$ 	& 	$41 \pm 7$ 	& 	$19 \pm 3$ 	& 	$14.1 \pm 1.6$		& 	$\sim 1893$						&	$1$		\\
HATLAS\,J001223.5$-$313242		&	SGP--213813 			& 	$3.47^{+0.40}_{-0.32}$	& 	$24 \pm 6$ 	& 	$35 \pm 8$ 	& 	$36 \pm 8$ 	&	$18 \pm 4$ 	& 	$13.9 \pm 0.7$ 		& 	$\sim 1866$						&	$1$		\\
HerMES\_J003352.2$-$452010	&	ELAISS1\_26 			& 						& 	$22 \pm 4$ 	&  	$34 \pm 5$ 	& 	$38 \pm 5$	&	$-$			& 	$13.5 \pm 0.6$		& 	$\sim 1812$						&	$1$		\\
HATLAS\,J083909.9+022718		&	G09--79552			& 	$3.59^{+0.34}_{-0.26}$	& 	$17 \pm 6$ 	& 	$38 \pm 8$ 	& 	$43 \pm 9$ 	& 	$17 \pm 4$ 	& 	$12.7 \pm 0.6$ 		& 	$\sim 1705$						&	$1$		\\
HATLAS\,J084113.6$-$004114		&	G09--59393			& 	$3.70^{+0.35}_{-0.26}$	& 	$24 \pm 7$ 	& 	$44 \pm 8$ 	& 	$47 \pm 9$ 	& 	$24 \pm 4$ 	& 	$12.4 \pm 0.4$ 		& 	$\sim 1665$						&	$2$		\\
HATLAS\,J222919.2$-$293731		&	SGP--385891			&	$3.70^{+0.29}_{-0.24}$	&	$13 \pm 8$	&	$46 \pm 10$	&	$60 \pm 11$	&	$21 \pm 4$	&	$11.1 \pm 0.7$		&	$\sim 1490$						&	$2$		\\
HATLAS\,J003131.1$-$293122		&	SGP--392029			&	$3.42^{+0.47}_{-0.32}$	& 	$18 \pm 7$	& 	$31 \pm 8$	& 	$35 \pm 8$	& 	$14 \pm 4$	& 	$10.8 \pm 0.8$		&	$\sim 1450$						&	$2$		\\
HATLAS\,J085156.0+020533		&	G09--80658			& 	$4.07^{+0.09}_{-0.72}$	& 	$18 \pm 6$ 	& 	$32 \pm 8$ 	& 	$40 \pm 9$ 	& 	$18 \pm 4$ 	& 	$10.7 \pm 0.7$ 		& 	$\sim 1437$	 					&	$2$		\\
HerMES\,J021743.9-030914		&	XMM\_15 				& 						& 	$25 \pm 5$ 	& 	$39 \pm 8$ 	& 	$47 \pm 7$ 	& 	$19 \pm 3$ 	& 	$ 9.4 \pm 0.5$		& 	$\sim 1262$						&	$1$		\\
HATLAS\,J084659.0$-$004219		&	G09--80620 			& 	$4.01^{+0.22}_{-0.78}$	& 	$14 \pm 5$ 	& 	$25 \pm 7$ 	& 	$28 \pm 8$ 	& 	$13 \pm 4$ 	& 	$ 8.4 \pm 0.7$		& 	$\sim 1128$						&	$2$		\\
HATLAS\,J231146.6$-$313518		&	SGP--386447 			& 	$4.89^{+0.78}_{-0.73}$	& 	$11 \pm 6$ 	& 	$34 \pm 6$ 	& 	$34 \pm 7$ 	& 	$34 \pm 8$ 	& 	$ 6.5 \pm 0.6$ 		& 	$\sim 873$						&	$1$		\\
HATLAS\,J001526.4$-$353738		&	SGP--135338			& 	$3.06^{+0.33}_{-0.26}$	& 	$33 \pm 7$ 	& 	$44 \pm 8$ 	& 	$53 \pm 9$ 	& 	$15 \pm 4$ 	& 	$ 6.1 \pm 0.4$ 		& 	$\sim 819$						&	$1$		\\


\hline
\hline
\tablenotetext{1}{Redshifts with error bars are photometric redshifts derived from {\it Herschel}+LABOCA/SCUBA-2 photometry, from \cite{Ivison2016ApJ...832...78I}.  Redshifts without error bars are unambiguous spectroscopic redshifts, taken from different works (see below).}
\tablenotetext{2}{From \cite{Asboth2016MNRAS.462.1989A}}
\tablenotetext{3}{From \cite{Fudamoto2017arXiv170708967F}}
\tablenotetext{4}{From \cite{Oteo2016ApJ...827...34O}}
\tablenotetext{5}{From \cite{Riechers2017arXiv170509660R}.}
\tablenotetext{6}{From \cite{Oteo2017arXiv170902809O}.}
\tablenotetext{7}{Flux densities obtained from the primary-beam corrected, tapered ALMA maps at $\sim 0.8''$ resolution.}
\tablenotetext{8}{No clear signatures of lensed features are seen in these ALMA high-resolution images, but the proximity to a near-IR source could indicate weak gravitational magnification.}
\tablenotetext{9}{Source with both lensed and unlensed components.}
\tablenotetext{10}{Number of components our ultrared DSFGs are resolved into.}
\end{tabular}
\end{table*}

\begin{figure}
\centering
\includegraphics[width=0.45\textwidth]{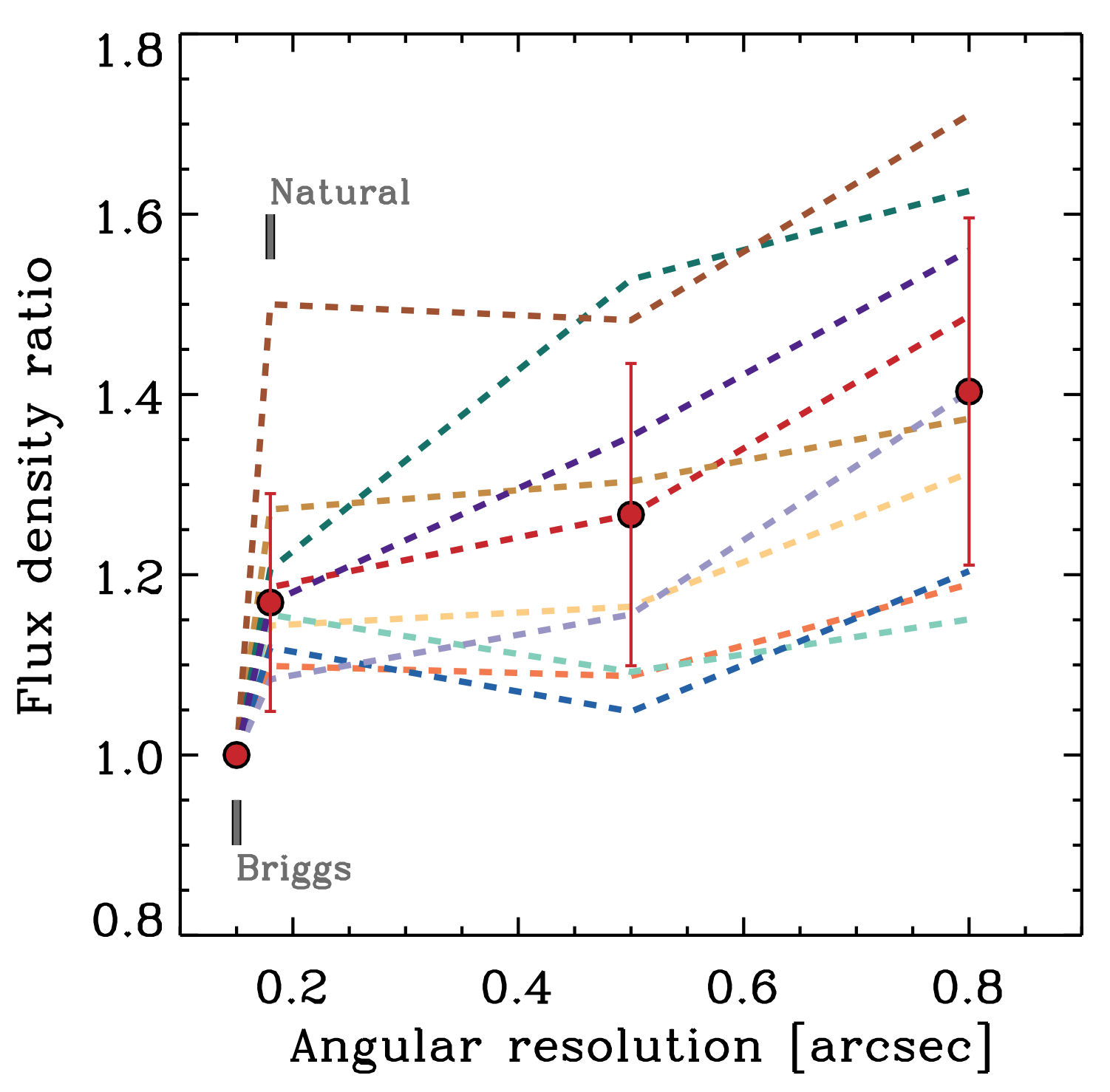} 
\caption{Flux density ratio for several of our unlensed ultrared DSFGs, for four different weighting schemes and $uv$ tapers (in order of increasing beam size): 1) Briggs weighting with ${\rm robust} = 0.5$, 2) natural weighting, 3) natural weighting and $\sim 0.5''$ tapering, and 4) natural weighting and $\sim 0.8''$ tapering.  The flux density ratios are measured with respect to Briggs weighting with robust=0.5.  The trend for individual sources is shown by the dashed curves, while red dots represent the average.  In most cases we see that natural weighting recovers a significant fraction of the flux recovered when tapering to $\sim 0.8''$. }
\vspace{5mm}
\label{flux_beamsize_imaging_fig}
\end{figure}

\begin{figure*}
\centering
\includegraphics[width=0.90\textwidth]{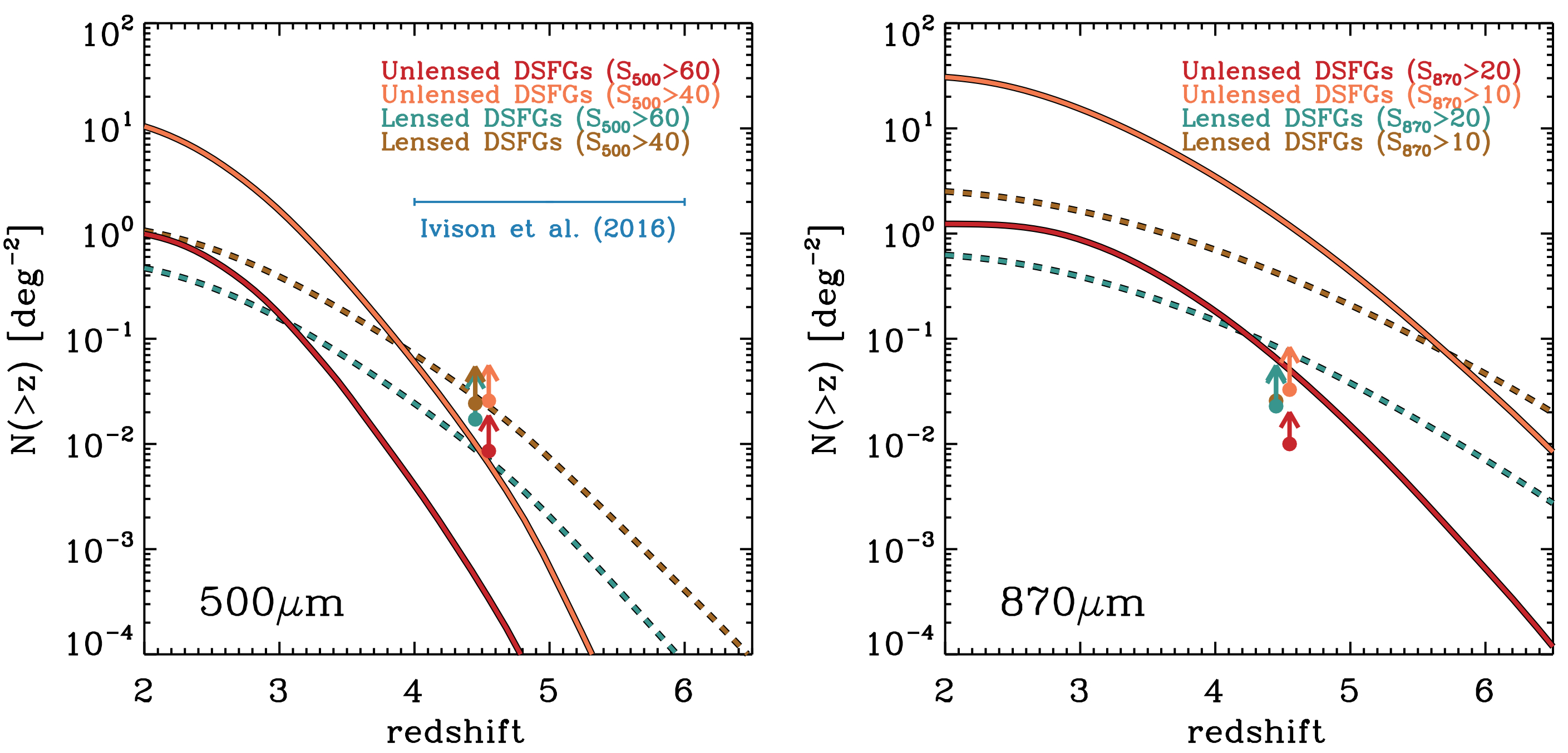}  
\caption{Expected number of lensed and unlensed DSFGs with $S_{\rm 500} > 40 \, {\rm mJy}$ and $S_{\rm 500} > 60 \, {\rm mJy}$ ({\it left}) and  $S_{\rm 870} > 20 \, {\rm mJy}$ and $S_{\rm 870} > 10 \, {\rm mJy}$ (where the flux densities have been obtained from the ALMA tapered maps at $0.8''$ resolution -- {\it right}) as a function of redshift according to the \cite{Cai2013ApJ...768...21C} models.  We consider here that the average redshift of the sample is $z=4.5$.  For the lensed DSFGs we have considered a maximum amplification factor of $\mu_{\rm max} = 15$ \citep{Negrello2017MNRAS.465.3558N}.  The vertical arrows represent lower limits found in our work.  Recall that only lower limits can be provided in this work because we study only the ultrared DSFGs observed with ALMA, not the full sample.  The arrows for lensed sources have been shifted to $z = 4.45$ and those for unlensed sources to $z = 4.55$, for clarity.  For reference, we include the number counts for the full sample of ultrared DSFGs in {\it H}-ATLAS taken from \citep{Ivison2016ApJ...832...78I}.  We see that models cannot reproduce the number of unlensed DSFGs  at 500\,$\mu$m in any of the two flux density ranges considered.  The model counts underestimation for unlensed sources is more significant at higher flux densities. The agreement with lensed sources is better, mostly in the flux density range $S_{500} > 40 \, {\rm mJy}$, although it may be that the 'true' number counts are much higher than the lower limits given here.  At 870\,$\mu$m we see that models might be able to reproduce the number of sources in our sample, although it is difficult to say how stringent our lower limits are.  In any case, we see a strong difference between the number counts at 500\,$\mu$m (measured from SPIRE) and at 870\,$\mu$m (measured from ALMA, after refinement with SCUBA-2 and LABOCA).  Possible reasons for this difference include flux boosting in SPIRE 500\,$\mu$m, which might affect the flux densities by tens of \%, or clustering in the SPIRE bands due to the large SPIRE beam. At the redshift of our ultrared DSFGs ($z = 4$--6) the models predict that a significant fraction of ultrared DSFGs are expected to be lensed, in agreement with our results.}
\vspace{5mm}
\label{number_counts_ultrared_cai_models_fig}
\end{figure*}

\section{ALMA high-resolution imaging at 870\,$\mu$m}
\label{section_ALMA_imaging_and_data}


Our sample of 44 ultrared DSFGs at $z \sim 4$--6 was observed between 2015 June and 2017 May when ALMA was in a relatively extended configuration.  Between 37 and 43 antennas were used, with distances from the center of the array of up to $1.6 \, {\rm km}$. The raw data were calibrated using the ALMA pipeline, then the calibrated visibilities were imaged using Briggs weighting with $\verb+robust+=0.5$, giving an average beam size of $0.12''$, or $\approx 800 \, {\rm pc}$ for an unlensed source at $z = 4.5$. The continuum r.m.s.\ sensitivity reached an average of 0.1\,mJy\,beam$^{-1}$.  Later on in the paper, we will refer to these as the `full-resolution maps', although slightly higher resolution would be achieved using uniform weighting which highlights more compact scales than Briggs with $\verb+robust+=0.5$.  We will also study the influence of different $uv$ weighting schemes on the recovered flux density, morphology and size of our sources, for which several {\sc robust} values will be considered along with uniform and natural weightings (the latter recovers more extended scales than Briggs weighting with $\verb+robust+=0.5$), and natural weighting with a Gaussian taper. 
 
At the frequency of our observations, the {\sc fwhm} of the ALMA primary beam is about $17''$ or $112 \, {\rm kpc}$ at $z = 4.5$.  Due to this relatively small field of view (FoV), it is plausible that we could fail to detect one or more companions to our ultrared DSFGs, especially for those sources where the LABOCA/SCUBA-2 emission appears extended.  We will go back to this issue in \S\ref{measuring_flux_densities_section}.  

\begin{figure*}[!t]
\centering
\includegraphics[width=0.90\textwidth]{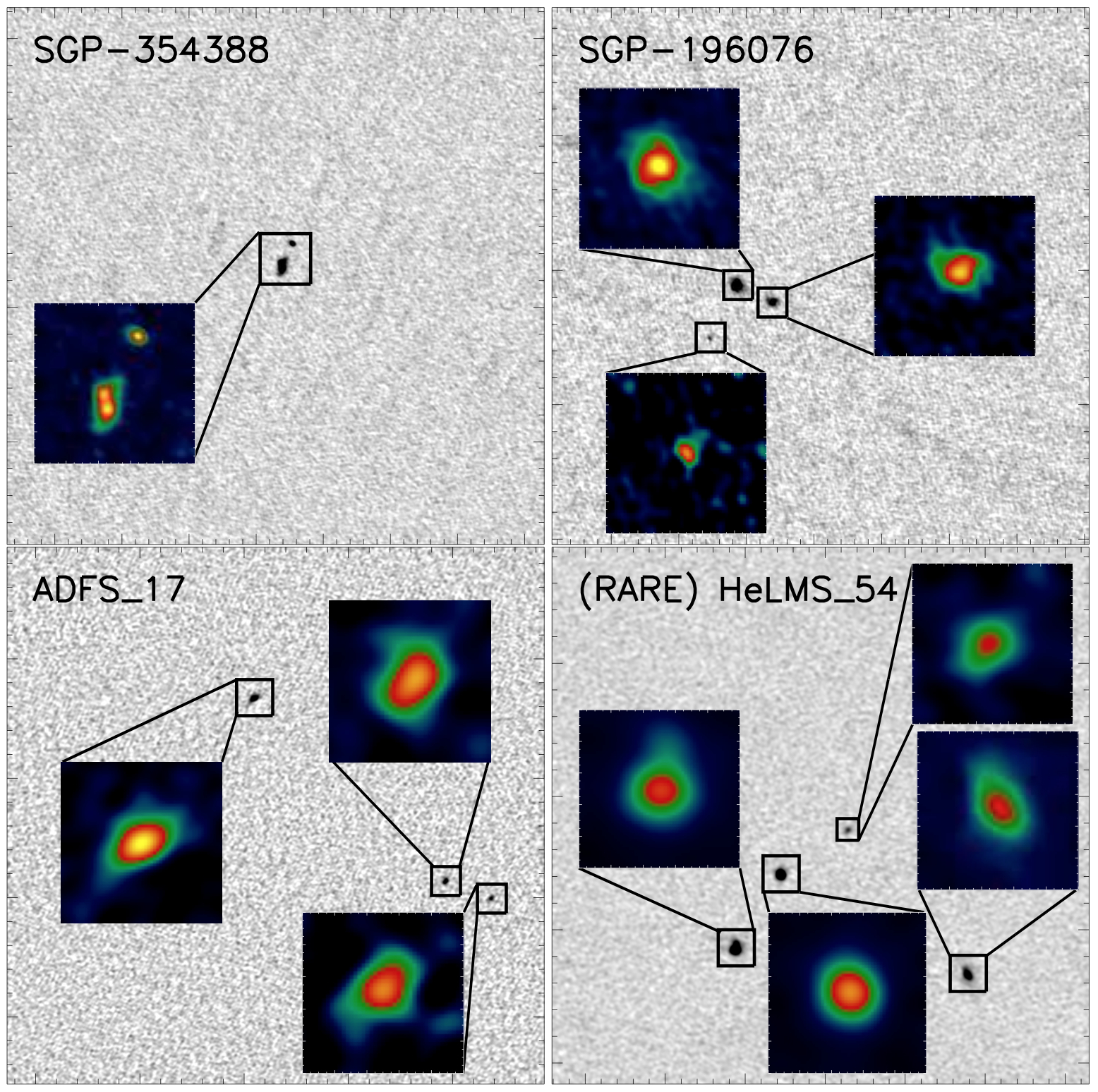} 
\caption{ALMA 870\,$\mu$m dust continuum imaging at $\sim 0.12''$ (or $\approx 800 \, {\rm pc}$) spatial resolution for four of our brightest ultrared DSFGs.  For clarity, the images shown here have not been corrected for the primary beam response. The images are $30''$ on each side and thus cover an area larger than the primary beam. Zoomed images are included to show the morphologies of the different sub-components into which our ultrared DSFGs are resolved. Images of the full sample are included in the Appendix.  We see a wide range of morphologies, from relatively extended interacting disks (e.g.\ SGP-196076 -- \citealt{Oteo2016ApJ...827...34O}) to compact starbursts (SGP-354388) or sources split into up to five components, such as (RARE) HeLMS\_54 (note that the faintest component is not shown in zoom, due to lack of space).
              }
\label{morphology_main_images_paper}
\end{figure*}

\subsection{Measuring flux densities}
\label{measuring_flux_densities_section}

The high-spatial resolution of our ALMA observations ($\sim 0.12''$ {\sc fwhm}) could mean that some of the 870\,$\mu$m flux density of our ultrared DSFGs is missed during imaging.  In this section, we explore how the recovered flux densities of our ultrared DSFGs vary as a function of the weighting schemes and $uv$ tapering used during imaging.  To accomplish this we also image the calibrated visibilities using natural weighting and natural weighting with two $uv$ tapers, giving two spatial resolutions: $\sim 0.5''$ and $\sim 0.8''$.

We show in Fig.~\ref{flux_beamsize_imaging_fig} the Briggs-to-natural flux density ratio (measured using the {\sc casa} task {\sc imfit} in the primary-beam corrected maps) of some of our ultrared DSFGs, randomly selected from the full sample (we do not include all sources in Fig.~\ref{flux_beamsize_imaging_fig} for the sake of clarity) as a function of the beam size of the maps.  All flux densities have been normalized to the flux density of the sources in the maps produced with Briggs weighting and {\sc robust} = 0.5.  We see in Fig.~\ref{flux_beamsize_imaging_fig} that the recovered flux densities increase with the beam size, as expected.  On average, natural weighting recovers most ($\sim 85\%$) of the flux density of our galaxies, assuming that the flux densities provided by the tapered maps at $\sim 0.8''$ resolution represent the true total flux (this is motivated by the fact that most sources are unresolved in the tapered maps at $\sim 0.8''$ resolution). On the other hand, the Briggs weighting misses $\sim 30\%$ of the total flux density.  These results are in agreement with \cite{Hodge2016ApJ...833..103H} for a sample of luminous, lower-redshift DSFGs observed at similar angular resolution. We note that while tapering to $\sim 0.8''$ resolution recovers most of the flux density of our sources, the angular resolution is then too poor to determine meaningful sizes.

We show in Table~\ref{table_flux_density_and_redshift} the primary-beam corrected flux density of our sources at 870\,$\mu$m, obtained from tapered maps at $0.8''$ resolution. The ALMA flux density at 870\,$\mu$m of the sources whose emission is resolved into several sub-components corresponds to the sum of the flux densities of all sub-components. The ALMA flux densities quoted in Table~\ref{table_flux_density_and_redshift} are measured at a similar frequency to the LABOCA/SCUBA-2 observations presented in \citet{Ivison2016ApJ...832...78I}. However, we see significant flux density differences between single-dish and ALMA observations in some of the sources. One of the most obvious cases is SGP-354388, for which LABOCA measured $S_{870} = 64 \pm 11 \, {\rm mJy}$ while the ALMA-derived flux density is $S_{B70} = 24.1 \pm 1.7 \, {\rm mJy}$. For this source, ultra-deep ALMA ${\rm 2 \, mm}$ observations reveal at least 10 DSFGs distributed over a $\sim 40''$ wide region \citep{Oteo2017arXiv170902809O}. While all these components are contributing to the LABOCA flux density (and the LABOCA emission appears extended due to the superposition of the 870\,$\mu$m emission of all these sources), only one of them (the brightest at ${\rm 2 \, mm}$) is covered by the ALMA FoV at 870\,$\mu$m. This suggests that some of the flux density losses in the ALMA maps might be due to multiple components lying outside of the ALMA band 7 FoV. There are also deep $3 \, {\rm mm}$ observations for SGP-196076, SGP-32338,  ADSF\,27, HELMS\_RED\_4, and HELMS\_RED\_31, and in all these cases the deep $3 \, {\rm mm}$ maps reveal that all sub-components contributing to the SCUBA-2/LABOCA flux density are within the ALMA band 7 primary beam.  For these sources, the SCUBA-2/LABOCA and ALMA flux densities are in good agreement, within the uncertainties. Due to the lack of $3 \, {\rm mm}$ imaging for most of the sources, we cannot explore further whether the differences in the flux densities between single-dish and ALMA observations are due to sources located outside of the ALMA FoV, but the few existing data do point in that direction.  We note that the SFRs quoted in Table~\ref{table_flux_density_and_redshift} have been obtained from the ALMA data (see \S\ref{section_SFR_most_luminous} for details), so if there are sources out of the ALMA FoV which contribute to the SPIRE flux densities and are at the same redshift, the SFRs of our targets would be even higher. The small FoV of our ALMA observations cannot explain those cases where the ALMA flux densities are higher than the LABOCA ones.  Examples include (RARE) HeLMS\_54 or (RARE) HeLMS\_42, which are among the brightest sources in our sample.  We have excluded the possibility that emission lines are contributing to the narrower ALMA band \citep{Smail2011MNRAS.414L..95S}.

\section{The most luminous starbursts in the early Universe}
\label{section_SFR_most_luminous}

The modest depth of the wide-field {\it H}-ATLAS and {\it Her}MES surveys means that our ultrared criterion selects among the brightest sources ever found at $z\sim 4$--6. Any sources confirmed to be unlensed would be the among the most luminous starbursts in the early Universe (some examples are shown in Fig.~\ref{morphology_main_images_paper}).  All unlensed DSFGs spectroscopically confirmed to lie at $z_{\rm spec} > 4$ so far had ${\rm SFR} \lesssim 3,000 \, M_\odot \, {\rm yr}^{-1}$, including GN20 at $z_{\rm spec} \sim 4.1$ \citep{Hodge2012ApJ...760...11H}, ALESS\,73.1 at $z_{\rm spec} \sim 4.76$ \citep{Coppin2009MNRAS.395.1905C}, AzTEC-1 at $z_{\rm spec} \sim 4.3$ \citep{Yun2015MNRAS.454.3485Y}, Vd-17871 at $z_{\rm spec} = 4.622$ \citep{Smolcic2015A&A...576A.127S}, or AzTEC-3 at $z_{\rm spec} \sim 5.3$ \citep{Riechers2014ApJ...796...84R}. Table~\ref{table_flux_density_and_redshift} lists the total SFRs of all the sources in our sample. These have been derived from their total IR luminosity, obtained by re-scaling the ALESS template \citep{Swinbank2014MNRAS.438.1267S} to their observed ALMA 870\,$\mu$m flux densities (in the tapered maps at $\sim 0.8''$ resolution, see \S\ref{measuring_flux_densities_section}).  We have used this method instead of fitting to the {\it Herschel} and LABOCA/SCUBA-2 photometry because the latter is more likely to be affected by confusion and by sources which outside of the ALMA FoV at 870\,$\mu$m.  This is justified because the ALESS template has been shown to provide a good representation of the FIR spectral energy distribution (SED) for most $z = 4$--6 ultrared DSFGs \citep{Ivison2016ApJ...832...78I}. We have used spectroscopic redshifts if available; otherwise, we have assumed the average photometric redshift of the sample $z \sim 4.5$. 

Using a fixed template to estimate the total SFRs of our galaxies adds an uncertainty.  For a source with $S_{870} = 20 \, {\rm mJy}$ at $z = 4$, the ALESS template (the one used in our work to measure SFRs) gives ${\rm SFR} \sim 2,700 \, M_\odot \, {\rm yr}^{-1}$.  Among the templates used in \cite{Ivison2016ApJ...832...78I}, which include a representative range of FIR SEDs, only the Cosmic Eyelash \citep{Swinbank2010Natur.464..733S,Ivison2010A&A...518L..35I} and \citet{Pearson2013MNRAS.435.2753P} templates give slightly lower SFRs: ${\rm SFR} \sim 2,300$ and $\sim 1,950 \, M_\odot \, {\rm yr}^{-1}$, respectively.  The template associated with Arp\,220 and the one reported by \citet{Pope2008ApJ...675.1171P} give higher SFRs, ${\rm SFR} \sim 4,900$ and $\sim 3,000 \, M_\odot \, {\rm yr}^{-1}$, respectively. The same happens for the templates associated with the lensed source G15.141 at $z = 4.24$ \citep{Lapi2011ApJ...742...24L,Cox2011ApJ...740...63C}, which gives ${\rm SFR} \sim 4,450 \, M_\odot \, {\rm yr}^{-1}$.  Although the uncertainties can be significant, we are therefore being conservative: the SFRs of our galaxies are not artificially high because of the chosen template but instead because they are genuinely extremely luminous.

The most luminous galaxy in our sample is (RARE) HeLMS\_54, with a total SFR of $ \sim 6,400 \, M_\odot \, {\rm yr}^{-1}$, but there are several other extreme, hyper-luminous starbursts, such as SGP-196076 (already studied in depth by \citealt{Oteo2016ApJ...827...34O}), SGP-317726, ADFS\_27 \citep{Riechers2017arXiv170509660R} and HeLMS\_RED\_68.

\begin{figure*}
\centering
\includegraphics[width=0.90\textwidth]{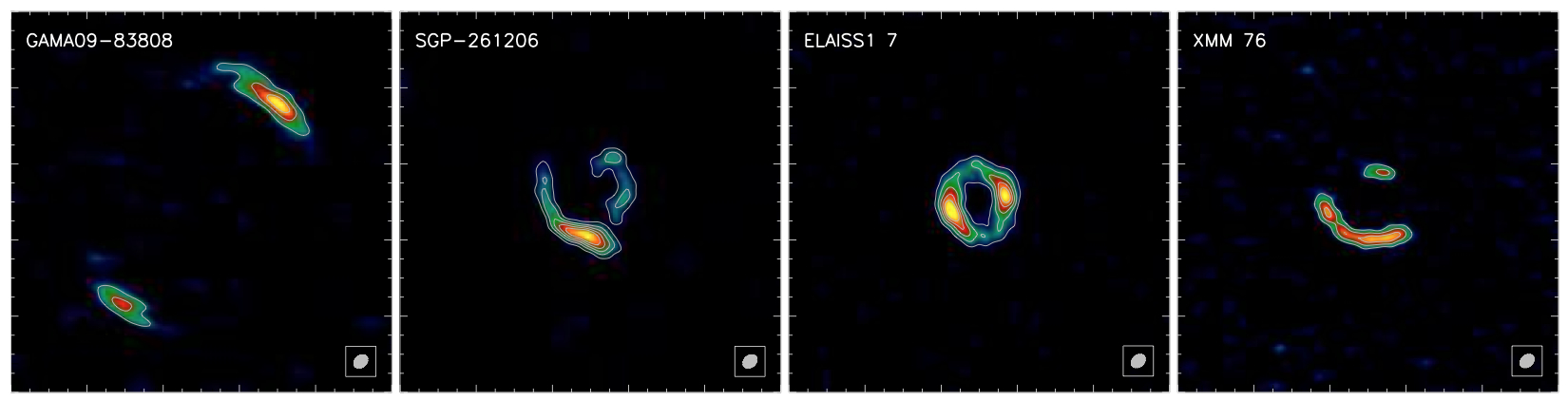}
\caption{ALMA 870\,$\mu$m dust continuum imaging of some of our lensed ultrared DSFGs at $z\sim 4$--6. All images are $4''$ on each side. Grey contours represent the dust continuum emission, from $5 \sigma$ in steps of $5 \sigma$.  We expect FIR-bright sources in our sample to be lensed \citep[see][and Fig.~\ref{number_counts_ultrared_cai_models_fig}]{Negrello2010Sci...330..800N} but we see here that some fainter galaxies are also lensed, such as XMM\,76.  From now on we will not consider lensed sources, since studying the morphology and size requires lens modeling that increases the uncertainties of our results.  Detailed analysis of the lensed sources will be presented elsewhere.}
\vspace{7mm}
\label{figure_lensed_URs}
\end{figure*}

\subsection{Comparison with models}

Fig.~\ref{number_counts_ultrared_cai_models_fig} shows model predictions for the number of unlensed DSFGs as a function of redshift at 500 and 870\,$\mu$m. In each case, two flux densities ranges are considered with the aim of comparing the brightest sources and the full sample: $S_{500} > 60 \, {\rm mJy}$ and $S_{500} > 40 \, {\rm mJy}$ for the 500\,$\mu$m flux densities and $S_{870} > 20 \, {\rm mJy}$ and $S_{870} > 10 \, {\rm mJy}$ for the 870\,$\mu$m flux densities (the latter measured from the ALMA maps, tapered to $\sim 0.8''$ resolution).  The associated number counts are $N (S_{500} > 60\,{\rm mJy}) > 8.6 \times 10^{-3} \, {\rm deg^{-2}}$, $N (S_{500} > 40\,{\rm mJy}) > 25.7 \times 10^{-3} \, {\rm deg^{-2}}$, $N (S_{870} > 20\,{\rm mJy}) > 10.0 \times 10^{-3} \, {\rm deg^{-2}}$ and $N (S_{870} > 10\,{\rm mJy}) > 32.9 \times 10^{-3} \, {\rm deg^{-2}}$. We highlight that in this work we can only provide lower limits on the number counts because not all ultrared DSFGs in our initial {\it Herschel} sample have been observed by ALMA.  The galaxy evolution models have been taken from \cite{Cai2013ApJ...768...21C} (see also \citealt{Negrello2017MNRAS.465.3558N}).  These models reproduce the SCUBA and ALMA $850 \, {\rm \mu m}$ number counts at the faint end fairly well \citep{Oteo2016ApJ...822...36O,Knudsen2008MNRAS.384.1611K} and also the SCUBA number counts \citep{Coppin2006MNRAS.372.1621C} at the bright end (it slightly over-predicts the latest SCUBA-2 number counts from \citealt{Geach2017MNRAS.465.1789G} at the bright end). The small disagreement with observations might be due to the fact that these models adopt a single, representative FIR SED (resembling that of the Cosmic Eyelash), and dust temperature and dust emissivity variations affect the number counts at FIR, submm and mm wavelengths. 

The \cite{Cai2013ApJ...768...21C} models at 500\,$\mu$m are not able to predict the number of unlensed ultrared DSFGs that we see in our sample, in either of the flux density ranges considered.  The models underestimation is more significant at higher flux density levels, where our sources are about $10 \times$ more abundant than predicted.  In principle, this might suggest that revision should be applied to models so that they can reproduce the number of observed sources.  However, the disagreement is less evident when comparing with the number counts at 870\,$\mu$m. In the right panel of Fig.~\ref{number_counts_ultrared_cai_models_fig} we see that the lower limits are still compatible with the models, although it is difficult to assess how stringent the lower limits really are.  In any case, it is evident that there is a significant difference when comparing our number of sources with models at 500\,$\mu$m (where the flux densities have been obtained from SPIRE observations) and at 870\,$\mu$m (where the flux densities have been obtained from our ALMA observations).

How can we explain the differences between 500 and 870\,$\mu$m?  First, our SPIRE flux densities  have not been corrected for flux boosting. As discussed in \cite{Ivison2016ApJ...832...78I}, this is because the correction factor is more uncertain than the correction itself. This lack of correction means that the SPIRE flux densities are likely over-estimated by tens of \% in some sources. Another issue potentially affecting the SPIRE flux densities is source clustering, due to the relatively large SPIRE beam, especially at $500\,{\rm \mu m}$ where flux densities could be contaminated by FIR emission from other IR-bright sources, close by.  These effects have clearly not affected the ultrared selection technique responsible for selecting our sample,  since most sources in our sample with spectroscopic redshifts have been confirmed to lie at $z_{\rm spec} \sim 4$--6.  This is likely because of the very significant sample refinement that was possible by imaging the ultrared DSFG candidates at higher spatial resolution and at longer wavelengths with SCUBA-2 and LABOCA \citep[see][]{Ivison2016ApJ...832...78I}.

The ALMA flux densities are not affected by either of these issues (although they might be under-estimated due to the relatively small FoV in band 7).  We therefore argue that the number counts at 870\,$\mu$m are a more reliable observable to compare with models.  In this way, the tension between models and observations might not be as significant as previously thought \citep{Asboth2016MNRAS.462.1989A}.  In order to more robustly test the models we require ALMA high-resolution observations of the full sample of ultrared DSFGs, preferably with small mosaics to cover more area, and we also need unambiguous spectroscopic redshifts to confirm that  our ultrared DSFGs lie at $z\sim 4$--6 and to confirm that the multiple sub-components into which our ultrared DSFGs are typically resolved are all at the same redshift and thus belong to the same starbursting group.

\begin{figure*}
\centering
\includegraphics[width=0.90\textwidth]{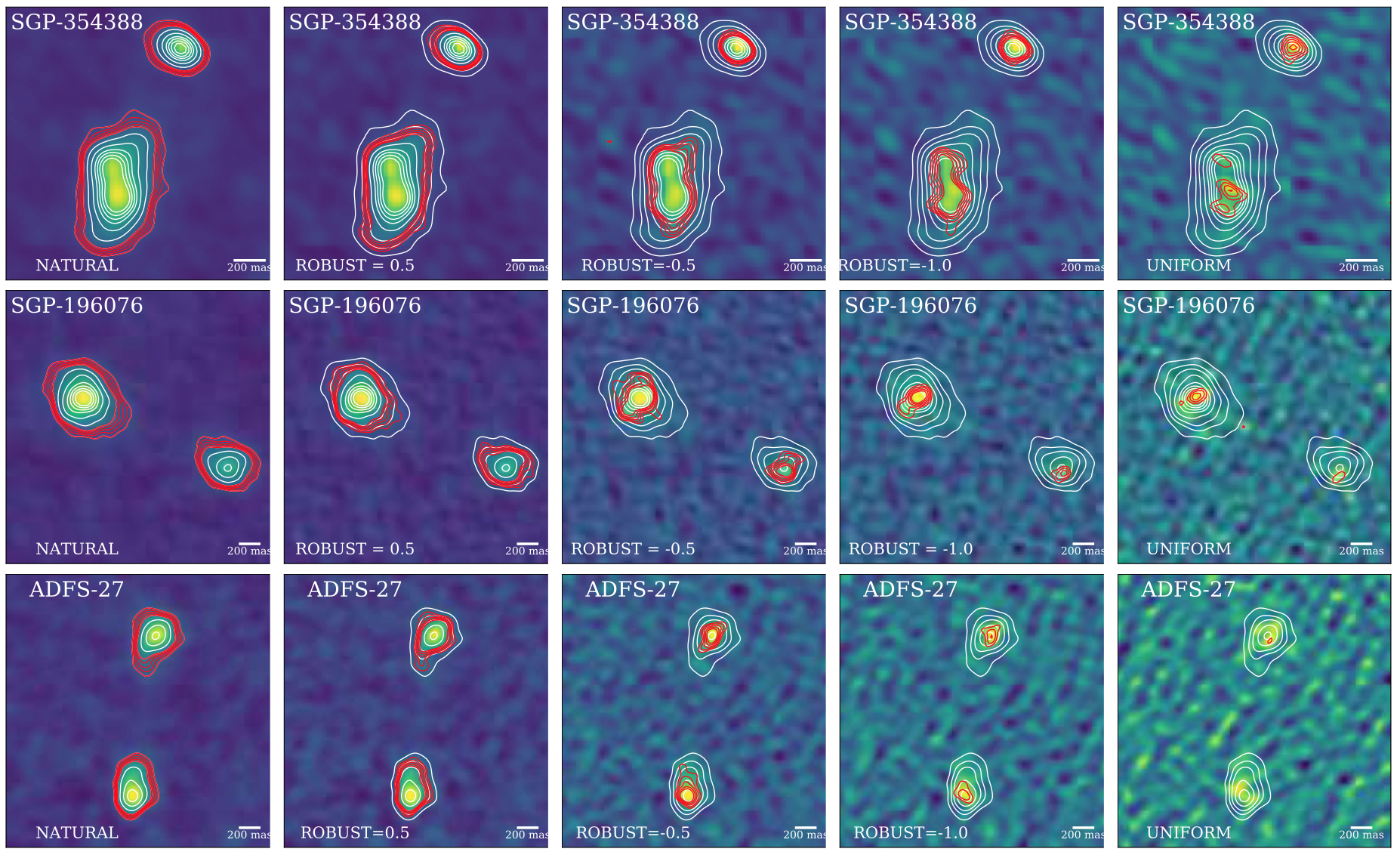}
\caption{Three of our brightest ultrared DSFGs -- each row corresponds to one source -- imaged using different weighting schemes. From left to right: natural, Briggs with different values of the robust parameter, and uniform.  These reveal their extended and compact emission to different degrees.  Natural weighing recovers more extended emission, while uniform weighting highlights the most compact emission.  Therefore, on each row, from left to right, we see from the most extended to the most compact features. On each row, white contours (starting from $4 \sigma$, in steps of $4 \sigma$) represent the emission in the map obtained with natural weighting.  Red contours (from 4 to $8 \sigma$ in steps of $1 \sigma$) represent the emission in each individual map.  We see that when using uniform weighting (which highlights the most compact emission) most of the emission disappears, confirming that a significant fraction of the dust emission in our ultrared DSFGs is extended relative to the ALMA synthesized beam used here.}
\vspace{5mm}
\label{flux_weighting_ALMA}
\end{figure*}

\section{Lensed ultrared DSFGs at $z \sim 4$--6}
\label{section_lenses}

The spatial resolution of our ALMA observations allows us to identify the lensed ultrared DSFGs in our sample by looking for signatures of lensed emission such as elongations, arcs and rings.  The high-resolution images of some lensed ultrared DSFGs in our sample are presented in Fig.~\ref{figure_lensed_URs} (the full sample is shown in Figures \ref{morphology_WIDE_map_1}, \ref{morphology_WIDE_map_2}, \ref{morphology_WIDE_map_3}, and \ref{morphology_WIDE_map_4}, except XMM\_30 and HeLMS\_RED\_4, which will be presented in a forthcoming paper) where lensing signatures are evident in all the sources. In addition to the lensing features, we have also explored possible lensing effects by using the available deep near-IR imaging in our fields: if there is a relatively small offset ($\lesssim 1.5-2''$) between the ALMA position of an ultrared DSFG and a near-IR source, then the ultrared DSFG is likely lensed by the near-IR source, which would likely lie at lower redshift than the ultrared DSFG. Doing this, we realized that even though there is no clear evidence of lensed signatures in the images of (RARE) HeLMS\_42 and HeLMS\_RED\_69, their proximity to a near-IR source may well indicate weak gravitational magnification \citep[see, e.g., galaxy T in][]{Ivison2013ApJ...772..137I}.  As noted by \cite{Fudamoto2017arXiv170708967F},  the lack of a near-IR counterpart at the depth of, for example, the VIKING survey \citep{Edge2013Msngr.154...32E}, which is the deepest available in some of our fields,  does not guarantee that a source is unlensed.  DSFGs gravitationally amplified by a galaxy cluster sometimes lack clear foreground counterparts \citep[e.g.][]{Nayyeri2017arXiv170101121N}.

We find 18 lensed ultrared DSFGs, representing 40\% of our ALMA sample.  The majority of the lensed sources are in the bright flux density regime, although others are fainter, such as XMM\,76.

Fig.~\ref{number_counts_ultrared_cai_models_fig} shows that models suggest a notable contribution of lensed sources in our flux density and redshift ranges.  At $S_{870} > 20 \, {\rm mJy}$, models predict that the number of lensed ultrared DSFGs is higher than the number of unlensed ones, at least at $z > 4.5$, while the number of lensed ultrared DSFGs is about $\sim 2-3\times$ lower than the number of unlensed at $S_{870} > 10 \, {\rm mJy}$.  Due to the different sources of incompleteness in our sample we cannot derive reliable fractions of lensed sources as a function of redshift or flux density, but we can trivially conclude that the fraction of lensed sources seen in our sample is not surprising according to models.


 
In this work we aim to measure accurate sizes and morphologies for our ultrared DSFGs.  For this reason we will exclude the lensed sources from our analysis, since the need for lens modeling will significantly affect the uncertainties of the results.  Detailed analysis of the lensed sources will be presented elsewhere.



\section{Morphology and physical scale of dust emission at $z > 4$}
\label{sec_morph_scale_dust}

We analyze in this section the morphologies and the physical scales of the dust emission in our sample of unlensed, ultrared DSFGs at $z \sim 4$--6. Note that the morphology of the dust emission in SGP-196076 and ADFS\,27 has been already analyzed using data from the ALMA project used in this work, by \cite{Oteo2016ApJ...827...34O} and \cite{Riechers2017arXiv170509660R}, respectively, but we include these sources here for the sake of completeness. 


\subsection{Morphologies}
\label{section_morph_UR}

The fine spatial resolution of our ALMA 870\,$\mu$m observations allows us to study the morphology of a large sample of DSFGs at $z\sim 4$--6 in unprecedented detail -- see Figs~\ref{morphology_main_images_paper}, \ref{morphology_WIDE_map_1}, \ref{morphology_WIDE_map_2}, \ref{morphology_WIDE_map_3}, and \ref{morphology_WIDE_map_4}. We see a diverse variety of morphologies: relatively extended smooth, disk-like shapes (SGP-196076); dust emission resolved into several interacting, compact components (SGP-354388); several isolated compact sources (ADFS\,17) and single, compact sources (XMM\,15).  The last column in Table~\ref{table_flux_density_and_redshift} gives the number of components our ultrared DSFGs are resolved into (we consider only $>5\sigma$ continuum detections, to avoid spurious sources -- \citealt{Oteo2016ApJ...822...36O}). The remarkable source multiplicity for HeLMS\,23, with its four bright components, is compatible with the findings for the classical SMG population at $z \sim 2.5$ \citep{Karim2013MNRAS.432....2K,Hodge2013ApJ...768...91H,Simpson2015ApJ...799...81S}. These multiple components do not necessarily have to lie at the same redshift and be physically related to each other, but having so many bright, unrelated sources within one ALMA FoV at 870\,$\mu$m is unlikely given the most recent ALMA number counts at 870\,$\mu$m \citep{Simpson2015ApJ...807..128S,Oteo2016ApJ...822...36O}

To further explore the presence of extended star formation in some of our ultrared DSFGs we have used different weighting schemes during the imaging and cleaning processes (some results are shown in Fig.~\ref{flux_weighting_ALMA}).  We see that when going from natural weighing (which recovers more extended emission) to uniform weighing (which highlights the most compact emission), most of the dust emission in our ultrared DSFGs disappears.  For example, when using uniform weighting, the dust emission is barely detected in the two interacting components of SGP-196076 (only the center of the brightest component remains detected) and ADFS\,27.  Therefore, we conclude that a significant part of the star formation in our sources is extended, and only a few cases show very compact emission, like in the northern component of SGP-354388. This result also suggests that caution should be taken when analyzing the morphology and size of the dust emission in high-redshift DSFGs, since these are dependent on the weighting schemes used to create the interferometric maps \citep[see also \S\ref{section_size_URs} and][]{Hodge2016ApJ...833..103H}.

There is no obvious trend between the morphology of our ultrared DSFGs and their SFRs.  The likely reason is that our sample contains only the most luminous DSFGs at $z = 4$--6; in order to study such trend, a less luminous control sample would be more informative. Actually, we find ultrared DSFGs which are resolved into multiple components both at our bright (see for example (RARE) HeLMS\_54) and faint end (see for example SGP-392029).

\begin{figure}
\centering
\includegraphics[width=0.49\textwidth]{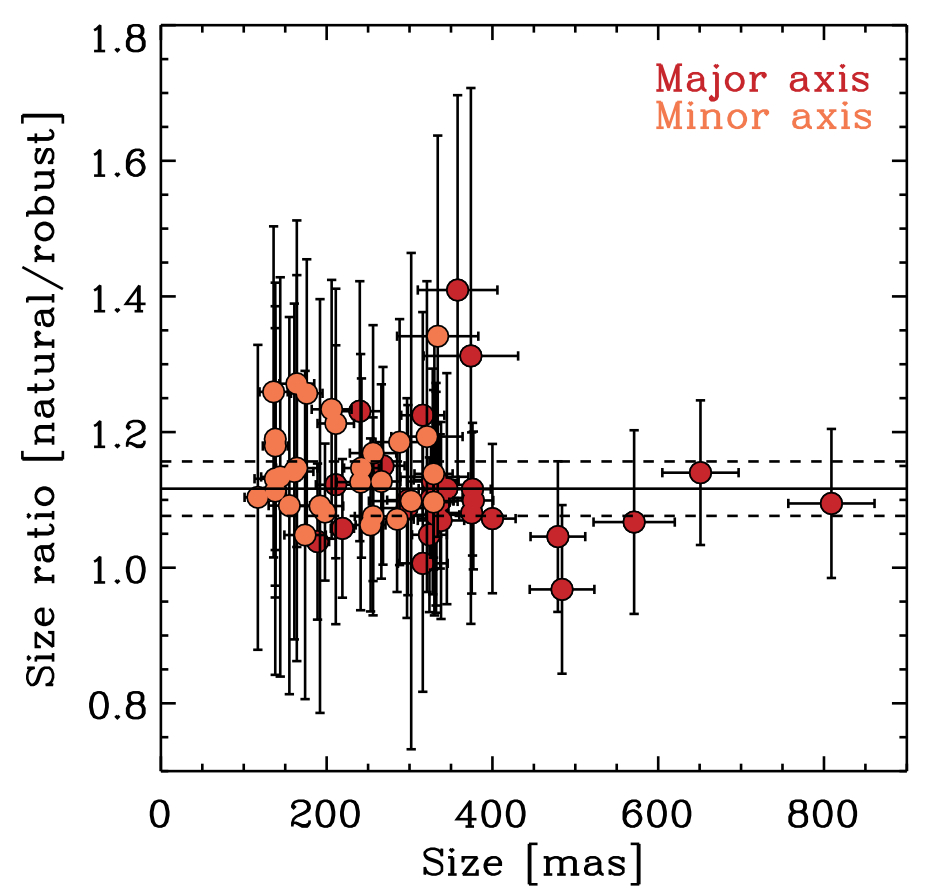} 
\caption{Ratio of the size of our ultrared DSFGs obtained with natural weighting and Briggs weighting with {\sc robust = 0.5} as a function of the ratio in the Briggs weighted maps with {\sc robust = 0.5} .  The size ratio for the major and minor axis are plotted separately, as indicated in the color legend.  We see that the sizes obtained using natural weighting are slightly higher than those obtained with Briggs weighing (as expected), by an average factor of $1.12 \times$. We recall that the flux densities of our sources, on average, are $1.17 \times$ times higher when using natural weighting with respect to Briggs weighting and {\sc robust = 0.5} (see Fig.~\ref{flux_beamsize_imaging_fig}).}
\vspace{5mm}
\label{figure_size_size_relation}
\end{figure}

\begin{table*}
\caption{\label{table_components_flux_and_area}Properties of the components seen in the unlensed, ultrared DSFGs}
\centering
\begin{tabular}{cccccc}
\hline
Source	&	$S_{\rm B7}$\tablenotemark{(a)} & $A$\tablenotemark{a} &  $A$\tablenotemark{a} & ${\rm SFR}$ & $\Sigma_{\rm SFR}$\\
 & [${\rm mJy}$] & $[{\rm mas \times mas}]$ & $[{\rm kpc \times kpc}]$ & $[M_\odot \, {\rm yr}^{-1}$] & [$M_\odot \, {\rm yr}^{-1} \, {\rm kpc}^{-2}]$ \\
\hline\hline

(RARE) HeLMS\_54.1	&	$14.7 \pm 0.8$			&	$286 \pm 17 \times 210 \pm 16$		&	$1.9 \pm 0.1 \times 1.4 \pm 0.1$	&	$\sim 1974$		&	$\sim 945$			\\
(RARE) HeLMS\_54.2	&	$9.9 \pm 0.2$			&	$191 \pm 9 \times 138 \pm 11$			&	$1.3 \pm 0.1 \times 0.9 \pm 0.1$	&	$\sim 1329$		&	$\sim 1447$			\\
(RARE) HeLMS\_54.3	&	$11.5 \pm 0.7$			&	$409 \pm 27 \times 204 \pm 19$		&	$2.7 \pm 0.2 \times 1.4 \pm 0.1$	&	$\sim 1544$		&	$\sim 520$			\\
(RARE) HeLMS\_54.4	&	$1.4 \pm 0.2$			&	--								&	--							&	$\sim 188$		&	--			\\
(RARE) HeLMS\_54.5	&	$1.1 \pm 0.3$			&	--								&	--							&	$\sim 148$		&	--			\\
SGP-196076\_1 		& 	$17.58 \pm 1.03$ 		& 	$337 \pm 25 \times 298 \pm 24 $ 		& 	$ 2.2 \pm 0.2 \times 2.0 \pm 0.2$ 	& 	$\sim 2370$ 		& 	$\sim 686$ 	\\
SGP-196076\_2 		&	$7.90 \pm 0.60$ 		&	$309 \pm 28 \times 238 \pm 24 $		& 	$ 2.0 \pm 0.2 \times 1.6 \pm 0.2$ 	& 	$\sim 1060$ 		& 	$\sim 384$ 	\\
SGP-196076\_3 		& 	$1.33 \pm 0.17$ 		& 	--								& 	-- 							& 	$\sim 180$ 		& 	--				\\
SGP-354388\_1 		& 	$9.64 \pm 0.33$ 		&	$255 \pm 10 \times 161	\pm 10 $ 		& 	$ 1.8 \pm 0.1 \times 1.1 \pm 0.1$ 	& 	$\sim 1301$ 		& 	$\sim 837 $\\
SGP-354388\_2 		& 	$3.61 \pm 0.25$ 		&	$170 \pm 17 \times 54 	\pm 47 $ 		& 	$ 1.2 \pm 0.2 \times 0.4 \pm 0.3$ 	& 	$\sim 487$ 		& 	$\sim 1015 $\\
SGP-354388\_3 		& 	$3.58 \pm 0.16$ 		&	$74 \pm 14 \times 65 \pm15 $ 			& 	$ 0.5 \pm 0.1 \times 0.5 \pm 0.1$ 	& 	$\sim 483$ 		& 	$\sim 1932 $\\
SGP-499646\_1		& 	$1.91 \pm 0.16$ 		&	$120 \pm 17	\times 58 \pm 34$ 		& 	$ 0.8 \pm 0.1 \times 1.5 \pm 0.2$ 	& 	$\sim 161$ 		& 	$\sim 170 $\\
SGP-32338\_1 		& 	$9.05 \pm 0.91$ 		&	$292 \pm 40 \times 268 \pm 39$ 		& 	$ 1.9 \pm 0.3 \times 1.8 \pm 0.3$ 	& 	$\sim 1219$ 		& 	$\sim 454 $\\
SGP-32338\_2 		& 	$2.10 \pm 0.25$ 		&	-- 								& 	-- 							& 	$\sim 283$ 		& 	--\\
ELAISS1\_40 			& 	$4.74 \pm 0.62$ 		&	$219 \pm 37 \times 168 \pm 34$ 		& 	$ 1.5 \pm 0.2 \times 1.1 \pm 0.2$ 	& 	$\sim 639$ 		& 	$\sim 493 $\\
SGP-386447 			&	$7.33 \pm 0.91$ 		&	-- 								& 	-- 							& 	$\sim 739$ 		& 	-- \\
SGP-93302 			& 	$9.95 \pm 0.90$ 		&	$260 \pm 31 \times 219 \pm 28$ 		& 	$ 1.7 \pm 0.2 \times 1.4 \pm 0.2$ 	& 	$\sim 1340$ 		& 	$\sim 717 $\\
SGP-317726 			& 	$4.74 \pm 0.55$ 		&	$191\pm 31 	\times 167 \pm 29$ 		& 	$ 1.3 \pm 0.2 \times 1.1 \pm 0.2$ 	& 	$\sim 640$ 		& 	$\sim 570 $\\
XMM\_15 				& 	$6.67 \pm 0.36$ 		&	$207\pm 15 	\times 116 \pm 13$ 		& 	$ 1.4 \pm 0.1 \times 0.8 \pm 0.1$ 	& 	$\sim 899$ 		& 	$\sim 1022 $\\
ADFS\,17\_1 			& 	$7.73 \pm 0.64$ 		& 	$253\pm 27 \times 109 \pm 18$ 		& 	$ 1.7 \pm 0.2 \times 0.7 \pm 0.1$ 	& 	$\sim 1042$ 		& 	$\sim 1115 $\\
ADFS\,17\_2 			& 	$5.29 \pm 0.49$ 		&	$199\pm 27 \times 104 \pm 30$ 		& 	$ 1.3 \pm 0.2 \times 0.7 \pm 0.2$ 	& 	$\sim 713$		& 	$\sim 998 $\\
ADFS\,17\_3 			& 	$4.97 \pm 0.68$ 		&	--								& 	-- 							& 	$\sim 670$ 		& 	--\\
ADFS\,27\_1 			& 	$10.17 \pm 0.83$ 		&	$271 \pm 27 \times 174 \pm 23$ 		& 	$ 1.8 \pm 0.2 \times 1.1 \pm 0.1$ 	& 	$\sim 1367$ 		& 	$\sim 879 $\\
ADFS\,27\_2 			& 	$9.83 \pm 0.95$ 		&	$317 \pm 34 \times 127 \pm 26$ 		& 	$ 2.1 \pm 0.2 \times 0.8 \pm 0.1$ 	& 	$\sim 1322$ 		& 	$\sim 1002 $\\
ADFS\,31\_1 			& 	$10.53 \pm 0.72$ 		&	$294\pm 23 \times 185 \pm 15$ 		& 	$ 1.9 \pm 0.2 \times 1.2 \pm 0.1$ 	& 	$\sim 1419$ 		& 	$\sim 793 $\\
ADFS\,31\_2 			& 	$2.53 \pm 0.30$ 		&	-- 								& 	--							& 	$\sim 341$ 		& 	-- \\
HeLMS\_182 			& 	$11.1 \pm 0.7$ 			& 	$247\pm 18 \times 122 \pm 12$ 		& 	$ 1.6 \pm 0.1 \times 0.8 \pm 0.1$	& 	$\sim 1492$ 		& 	$\sim 1485 $\\
ELAISS1\_18\_1 		& 	$11.2 \pm 0.9$ 			&	$471 \pm 41 \times 233 \pm 24$ 		& 	$ 3.1 \pm 0.3 \times 1.5 \pm 0.2$ 	& 	$\sim 1506$ 		& 	$\sim 413 $\\
ELAISS1\_18\_2 		& 	$4.0 \pm 0.4$ 			&	$205 \pm 28 \times 184 \pm 30$ 		& 	$ 1.4 \pm 0.2 \times 1.2 \pm 0.2$ 	& 	$\sim 538$ 		& 	$\sim 408 $\\
ELAISS1\_26 			& 	$8.63 \pm 0.58$ 		&	$233 \pm 20 \times 167 \pm 17$ 		& 	$ 1.5 \pm 0.1 \times 1.1 \pm 0.1$ 	& 	$\sim 1160$ 		& 	$\sim 896 $\\
SGP-72464			&	$13.23 \pm 0.92$		&	$347 \pm 28 \times 236 \pm 22$		& 	$ 2.3 \pm 0.2 \times 1.6 \pm 0.1$	&	$\sim 1779 $		&	$\sim 616$			\\
SGP-392029.1			&	$5.15 \pm 0.28$		&	$182 \pm 15 \times 124 \pm 13$		& 	$ 1.2 \pm 0.1 \times 0.8 \pm 0.1$	&	$\sim 692 $		&	$\sim 918$			\\			
SGP-392029.2			&	$3.11 \pm 0.24$		&	$188 \pm 27 \times 166 \pm 30$		& 	$ 1.2 \pm 0.2 \times 1.1 \pm 0.2$	&	$\sim 418 $		&	$\sim 403$			\\			
SGP-135338			&	$5.06 \pm 0.28$		&	$307 \pm 20 \times 106 \pm 16$		& 	$ 2.0 \pm 0.1 \times 0.7 \pm 0.1$	&	$\sim 680 $		&	$\sim 619$			\\
SGP-213813			&	$9.84 \pm 0.87$		&	$535 \pm 50 \times 219 \pm 26$		& 	$ 3.5 \pm 0.3 \times 1.4 \pm 0.2$	&	$\sim 1323 $		&	$\sim 344$			\\
G09-80620.1			&	$5.08 \pm 0.97$		&		--							&	--							&	$\sim 683 $		&	--			\\
G09-80620.2			&	$1.45 \pm 0.29$		&		--							&	--							&	$\sim 195 $		&	--			\\
G09-80658.1			&	$4.08 \pm 0.40$		&	$309 \pm 36 \times 143 \pm 28$		& 	$ 2.0 \pm 0.2 \times 0.9 \pm 0.2$	&	$\sim 549 $		&	$\sim 389$			\\
G09-80658.2			&	$1.35 \pm 0.13$		&	$139 \pm 35 \times 53 \pm 26 $		& 	$ 0.9 \pm 0.2 \times 0.3 \pm 0.2$	&	$\sim 181 $		&	$\sim 854$		\\
G09-79552			&	$10.09 \pm 0.85$		&	$273 \pm 29 \times 214 \pm 25$		& 	$ 1.8 \pm 0.2 \times 1.4 \pm 0.2$	&	$\sim 1357 $		&	$\sim 686$			\\
G09-59393.1			&	$7.33 \pm 0.48$		&	$373 \pm 28 \times 117 \pm 16$		& 	$ 2.5 \pm 0.2 \times 0.8 \pm 0.1$	&	$\sim 985 $		&	$\sim 627$			\\
G09-59393.2			&	$3.25 \pm 0.64$		&	--								&		--						&	$\sim 437 $		&	--			\\
HELMS\_RED\_68.1 	&	$6.06 \pm 0.54$		&	$500 \pm 50 \times 141 \pm 23$		& 	$3.3 \pm 0.3 \times 0.9 \pm 0.1$	&	$\sim 814 $		&	$\sim 349$			\\
HELMS\_RED\_68.2 	&	$16.5 \pm 2.2$			&	$314 \pm 51 \times 176 \pm 42$		& 	$2.1 \pm 0.3 \times 1.2 \pm 0.2$	&	$\sim 2218 $		&	$\sim 1121$			\\
HELMS\_RED\_270.1 	&	$7.94 \pm 0.61$		&	$458 \pm 37 \times 238 \pm 24$		&	$3.0 \pm 0.2 \times 1.6 \pm 0.2$ 	&	$\sim 1067 $		&	$\sim 283$			\\
HELMS\_RED\_270.2 	&	$5.73 \pm 0.51$		&	$291 \pm 32 \times 242 \pm 28$		& 	$1.9 \pm 0.2 \times 1.6 \pm 0.2$	&	$\sim 770 $		&	$\sim 323$			\\
SGP-385891.1 		&	$5.9 \pm 0.3$			&	$358 \pm 22 \times 134 \pm 15$		&	$2.4	\pm 0.1 \times 0.9 \pm 0.1$	&	$\sim 579$		&	$\sim 341$			\\
SGP-385891.2 		&	$1.5 \pm 0.2$			&	--								&	--							&	$\sim 147$		&	--			\\

\hline
\hline
\tablenotetext{1}{Measured in the primary-beam-corrected maps}

\end{tabular}
\tablenotetext{1}{The reported values are ${\rm FWHM_{major} \times FWHM_{minor}}$, where ${\rm FWHM_{major}}$ and ${\rm FWHM_{minor}}$ are obtained from a two-dimensional elliptical Gaussian profile fit to the observed emission. We only report the size of sources detected at $> 10\sigma$ at 870\,$\mu$m before primary-beam correction. In order to derive the physical size in kpc for sources without spectroscopic redshift we have assumed $z = 4.5$.}
\end{table*}

\subsection{Sizes}
\label{section_size_URs}

Before discussing the sizes of our ultrared DSFGs, we compare measurements obtained with Briggs weighting and {\sc robust = 0.5} and with natural weighting. This is shown in Fig.~\ref{figure_size_size_relation}. For each map and for each source, we have measured the size of the different components of our ultrared DSFGs by fitting their dust continuum images with 2D elliptical Gaussian profiles. This has been done with the {\sc casa} task, {\sc imfit}, as in previous work \citep{Simpson2015ApJ...799...81S}. On each map, we first fit the dust emission of the brightest component. Then, we fit the dust emission of the brightest component in the residual map (if any) from the previous fit. This is repeated until there are no remaining detections at $> 5 \sigma$ in the last residual map. It can be seen in Fig.~\ref{figure_size_size_relation} that the measured deconvolved sizes of our ultrared DSFGs is higher in the maps obtained with natural weighting with respect to those in the maps obtained with Briggs weighting and {\sc robust = 0.5}, as expected, by a factor of $1.12 \times$.  The change in size runs in parallel to the change in flux density reported in \S\ref{measuring_flux_densities_section}.  From now on, we report sizes by using the maps obtained with Briggs weighting and {\sc robust = 0.5}, bearing in mind that flux densities are $1.17 \times$ times higher and sizes are $1.12\times$ times higher in the natural weighted maps.

Table~\ref{table_components_flux_and_area} quotes the flux densities (obtained from the primary-beam-corrected maps) and sizes (de-convolved from the beam) of the components comprising each ultrared DSFG, in on-sky and physical units.  We report the flux densities of all components detected at $> 5 \sigma$, but report the size only of components detected at $> 10 \sigma$ because simulations have demonstrated that sizes of sources detected at $< 10 \sigma$ are not reliable \citep[e.g.][]{Simpson2015ApJ...799...81S,Hodge2016ApJ...833..103H}. In order to determine the size of each component in physical units (and also to determine luminosity distances and thence SFR -- see later) we use spectroscopic redshifts when available.  For sources without spectroscopic redshift, we have assumed the average photometric redshift of the sample, $z_{\rm phot} \sim 4.5$.  Assuming that our ultrared DSFGs lie $z \sim 4.5$ introduces extra uncertainty on the size determination in physical units, because for the assumed cosmology the angular scale is $7.3 \, {\rm kpc} / ''$ at $z = 3.5$ and $5.7 \, {\rm kpc} / ''$ at $z = 6.0$.


The SFR of each component (shown in Table~\ref{table_components_flux_and_area}) is calculated from its observed flux density assuming that the ALESS template provides a good representation of its FIR/submm SED (see \S\ref{section_SFR_most_luminous} for a discussion on the uncertainties related to this assumption).  It can be seen that the components of each ultrared DSFG are luminous starbursts, with ${\rm SFR} > 100 \, M_\odot \, {\rm yr}^{-1}$, sometimes as high as $\sim 2,400 \, M_\odot \, {\rm yr}^{-1}$.

\begin{figure}
\centering
\includegraphics[width=0.49\textwidth]{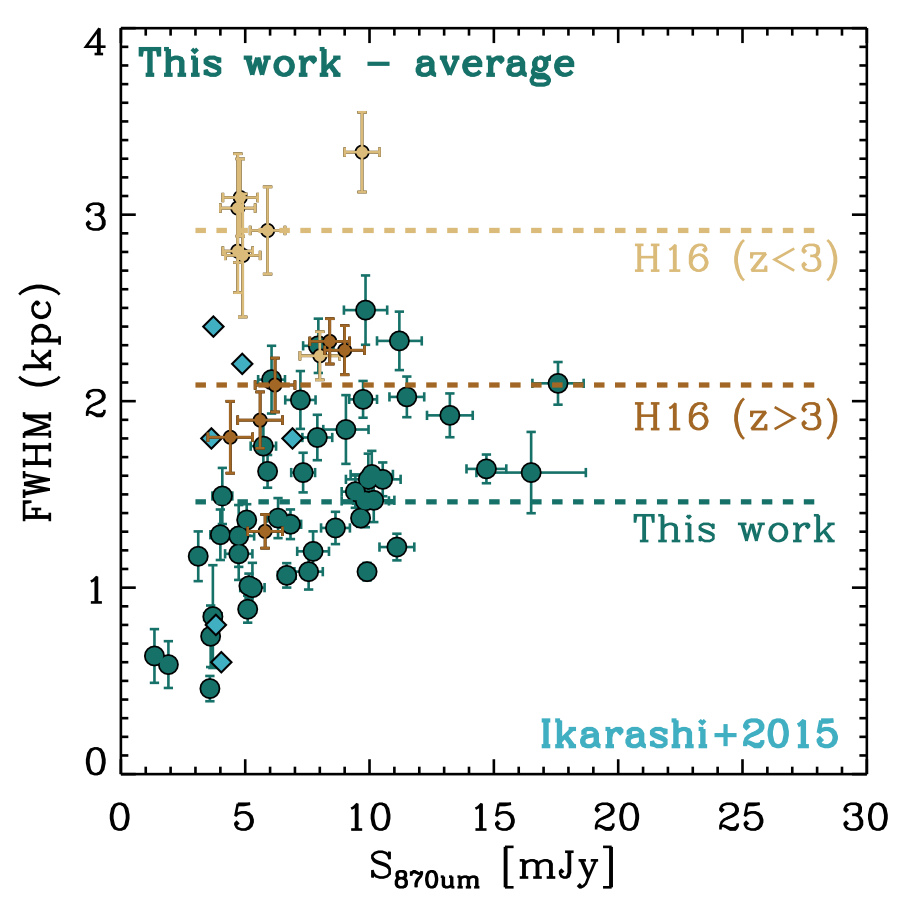} 
\caption{Size of our ultrared DSFGs as a function of their primary-beam-corrected flux densities at 870\,$\mu$m. We include in this figure the average of the major and minor axes, each calculated from a 2D Gaussian fit to the dust continuum emission.  The values reported correspond to the {\sc fwhm} of the Gaussian fits.  For comparison we also represent the size of the SMGs studied in \cite{Ikarashi2015ApJ...810..133I} and \cite{Hodge2016ApJ...833..103H}.  We distinguish between sources at $z_{\rm spec} < 3.0$ and $z_{\rm spec} > 3.0$ in the \cite{Hodge2016ApJ...833..103H} sample. The flux densities in \cite{Ikarashi2015ApJ...810..133I} were derived at $\sim 1.1\,{\rm mm}$ and have been converted to flux densities at $\sim 870 \, {\rm \mu m}$ by multiplying them by $\times 1.5$.  The average values found in \cite{Hodge2016ApJ...833..103H} and this work are indicated with the horizontal dashed lines. We see that DSFGs at higher redshifts tend to have smaller sizes, on average.}
\vspace{5mm}
\label{flux_rad_figure}
\end{figure}

Fig.~\ref{flux_rad_figure} shows the physical size of our ultrared DSFGs as a function of their observed, primary-beam-corrected flux densities at 870\,$\mu$m. We show the size ({\sc fwhm}) of the major axis, of the minor axis, and the average of the major and minor axes. We see evidence that brighter components are larger than smaller components. However, care should be taken when interpreting the size--flux relation: the faintest galaxies might have smaller sizes because the observations are not deep enough to detect any faint extended emission. However, any bright ($S_{870} > 15 \, {\rm mJy}$) and relatively compact ($\lesssim 1.5\,{\rm kpc}$) galaxies should have been found.  They are not present in our sample. We see a wide range of sizes for a fixed flux density, especially in the flux density range $5 < S_{870}\,[{\rm mJy}] < 10$, with an average size for our ultrared DSFGs, $\theta_{\rm FWHM}=1.46 \pm 0.41 \, {\rm kpc}$.

We also show in Fig.~\ref{flux_rad_figure} the sizes of luminous DSFGs at $z \sim 2.5$ selected from the ALESS survey \citep{Karim2013MNRAS.432....2K,Hodge2013ApJ...768...91H} derived by \cite{Hodge2016ApJ...833..103H} with dust continuum observations at a spatial resolution (in sky units) comparable to ours. Note that we exclude from the \cite{Hodge2016ApJ...833..103H} sample one source without a spectroscopic redshift and another source with $z_{\rm spec} < 1$.  Notably, the average size of the DSFGs in \cite{Hodge2016ApJ...833..103H} is $2\times$ larger than the average value found in this work. We have divided the \cite{Hodge2016ApJ...833..103H} sample into two redshift ranges: sources whose redshifts are greater than and less than $z_{\rm spec} = 3.0$ and it can be seen that the dust emission for DSFGs at lower redshift is more extended than for those at higher redshifts. \cite{Ikarashi2015ApJ...810..133I} reported the dust continuum sizes for a sample of AzTEC-selected SMGs at $z_{\rm phot} > 2.8$, although they are likely at lower redshift than our ultrared DSFGs given their bluer SPIRE colors. Despite the differences, the sizes reported in \cite{Ikarashi2015ApJ...810..133I} are similar to those for our ultrared DSFGs. In this comparison we have only used the six sources in \cite{Ikarashi2015ApJ...810..133I} with observations at a resolution similar to that in our work, since the resolution for the other galaxies is too low ($\sim 0.7''$) to allow a fair comparison. Our derived sizes are also smaller than the size of DSFGs at $z \sim 1$--3 measured via high-resolution radio continuum imaging \citep{Biggs2008MNRAS.385..893B} or in high-$J$ CO emission lines \citep{Tacconi2006ApJ...640..228T}. 

The combination of all previous results show that DSFGs tend to be smaller in the early Universe, $z = 4$--6, than at `cosmic noon' -- the peak of cosmic star formation, $z \sim 2$--3. 

\subsection{SFR surface density}

In this section we discuss the SFR surface density ($\Sigma_{\rm SFR}$) of our galaxies, which have been calculated as: $\Sigma_{\rm SFR} = {\rm SFR} / A_{\rm dust}$, where the area of the dust emission, $A_{\rm dust}$, has been obtained from $\pi \times R_{\rm a} \times R_{\rm b}$, where $R_{\rm a}$ and $R_{\rm b}$ are the semi-major and semi-minor axis, respectively, calculated as $R_{\rm a} = 0.5 \times {\rm FWHM}_{\rm major}$ and $R_{\rm b} = 0.5 \times {\rm FWHM}_{\rm minor}$. This is the same definition used in \cite{Oteo2016ApJ...827...34O}, \cite{Riechers2014ApJ...796...84R} and several other works analyzing the properties of dust emission in luminous starbursts at high redshift.  However, we note that some other work, e.g.\ \citet{Simpson2015ApJ...799...81S,Oteo2017ApJ...837..182O}, used the definition $\Sigma_{\rm SFR}^{'} = 0.5 \times {\rm SFR} / A_{\rm dust}$ to take into account that {\sc fwhm} measures the size where half of the total SFR is taking place. The difference between the two definitions is relevant when comparing different samples in the literature.

We find a noticeable range of $\Sigma_{\rm SFR}$ in our ultrared DSFGs, from $\sim 150$ to $\sim 2,000 \, M_\odot \, {\rm yr}^{-1} \, {\rm kpc}^{-2}$ (see values in Table~\ref{table_components_flux_and_area}). Remarkably, 14 of our unlensed ultrared DSFGs have $\Sigma_{\rm SFR} > 800 \, M_\odot \, {\rm yr}^{-1} \, {\rm kpc}^{-2}$ and are therefore at -- or are close to -- the Eddington limit \citep{Thompson2005ApJ...630..167T}. We have already discussed the Eddington-limited nature of SGP-196076\_1 in \cite{Oteo2016ApJ...827...34O}. Despite its high $\Sigma_{\rm SFR}$, there is no sign of outflows in the CO or [C\,{\sc ii}] emission, and the ${\rm OH\,163 \, \mu m}$ transition (a tracer of molecular outflows -- \citealt{Spoon2013ApJ...775..127S}) is weak. SGP-354388 and ADFS\,17 are interesting cases, since two of their interacting components are Eddington-limited starbursts.

It should be noted that determining $\Sigma_{\rm SFR}$ accurately for DSFGs requires us to measure the extent of their dust emission accurately and, thus, on the availability of observations with sufficient resolution to resolve individual star-forming clumps. Some of the apparently smooth disks could be resolved into different star-forming clumps if observed at higher resolution. This was, for example, the case for ALMACAL-1 and ALMACAL-2 \citep{Oteo2017ApJ...837..182O}, the two brightest SMGs found in the ALMACAL survey \citep{Oteo2016ApJ...822...36O}. Using high spatial resolution ($\sim 0.4''$) data we derived an SFR surface density of $\sim 165 \, M_\odot \, {\rm yr}^{-1} \, {\rm kpc}^{-2}$ in ALMACAL-1. However, ultra-high spatial resolution (beam size $\sim 20 \, {\rm mas}$) observations revealed that the sources are actually extremely compact and have enormous SFRs, meaning the SFR surface density is as high as $\sim 6,000 \, M_\odot \, {\rm yr}^{-1} \, {\rm kpc}^{-2}$. 


\subsection{From ultrared DSFGs to local ultra--massive ellipticals}

Having confirmed the existence of a population of DSFGs with extreme IR luminosities and SFRs at $z \sim 4$-- 6, we now explore the possible future evolution of these systems. 

Given the relatively wide range of morphologies and spatial extents of our ultrared DSFGs (see \S\ref{section_morph_UR}), the interpretation is challenging for many sources. There are ultrared DSFGs which are resolved into different sub-components separated by several arcsec (ADFS\,17, ADFS\,31, or SGP-93302 -- see Figs.~\ref{morphology_WIDE_map_1} to \ref{morphology_WIDE_map_2}). Without spectroscopic confirmation it is not possible to know whether all these galaxies are at the same redshift and are, therefore, physically associated.  On the other hand, there are sources that -- due to their close proximity -- are likely the components of a merger of several dusty galaxies. Examples include SGP-196076 (where the interaction has been confirmed via detection of CO and [C\,{\sc ii}] in two of the galaxies -- \citealt{Oteo2016ApJ...827...34O}), SGP-354388, ADFS\,27 and ELASISS1\,18. The physical relation between close components is further supported by the submm number counts \citep[see for example][]{Simpson2015ApJ...807..128S,Oteo2016ApJ...822...36O}, since the probability of finding such bright sources so close to each other is very low.

In order to study the likely evolution of our ultrared DSFGs we would ideally need $^{12}$CO(1--0) or [C\,{\sc i}](1--0) detections to measure their molecular gas mass which, in combination with their total SFRs, provide estimates of their gas-depletion times (always assuming a standard IMF, cf.\ \citealt{Romano2017MNRAS.470..401R}, who claim $L_{\rm IR}$-derived SFRs in starbursts might be up to $5\times$ too high).  In \cite{Oteo2016ApJ...827...34O} we estimated the gas-depletion time for one of our sources with mid-$J$ CO detections and concluded that -- in the absence of an external molecular gas supply -- this source would likely become a massive elliptical by $z \sim 3$.  The same would happen to ADSF\_27 according to the gas-depletion time derived by \cite{Riechers2017arXiv170509660R}.  Our galaxies lack of CO data, for the most part, but we can estimate the gas-depletion time using the relatively tight relation between the dust continuum luminosity at rest-frame $850 \, {\rm \mu m}$ and the $^{12}$CO(1--0) luminosity \citep[e.g.][]{Scoville2016ApJ...820...83S,Hughes2017MNRAS.468L.103H,Oteo2017arXiv170705329O} with the caution that this relation has not been tested for galaxies in our redshift and luminosity ranges. We have estimated the rest-frame $850 \, {\rm \mu m}$ luminosity of our sources by assuming the ALESS template, providing some consistency with the procedure we followed to estimate their SFR.  The average flux density of our sources at 870\,$\mu$m is $7.7 \, {\rm mJy}$.  Assuming that the average redshift of our sources is $z = 4.5$, the average total SFR would then be ${\rm SFR \sim 760 \, M_\odot \, {\rm yr}^{-1}}$ and the molecular gas mass $M_{\rm gas} \sim 3.2 \times 10^{11} M_\odot$.  This implies a gas-depletion time of $420 \, {\rm Myr}$ and that all the molecular gas available for star formation will be consumed by $z \sim 3$.  Therefore, if there is no external gas supply, our ultrared DSFGs will evolve into massive elliptical-like galaxies at $z \sim 3$--3.5, with stellar masses of at least $M_{\rm stars} = 3.2 \times 10^{11} \, M_\odot$.

Extensive work on the analysis of the stellar populations in local elliptical galaxies has shown that the more massive a galaxy is, the earlier it should have been formed. In particular, most massive galaxies with $M_{\rm star} > 5 \times 10^{11} M_\odot$ in the local Universe formed most of their stars at $z > 2$ in a relatively fast and intense burst of star formation, which could range in duration between $\sim 1\,{\rm Gyr}$ to a few million years for the most massive galaxies \citep{Thomas2010MNRAS.404.1775T}. Along the same lines, the existence of a population of massive, red-and-dead galaxies at $z \sim 2$ implies an early formation epoch \citep{Krogager2014ApJ...797...17K}, a phase which is consistent with the extreme SFRs found in our ultrared DSFGs \citep{Toft2014ApJ...782...68T}. Taking together, our results suggest that ultrared DSFGs evolve into the most massive galaxies at $z \sim 3$, which are the progenitors of the already quiescent population at $z \sim 2$ which, in turn, are the progenitors of local ultra-massive galaxies. This scenario is similar to the one proposed by \cite{Ikarashi2015ApJ...810..133I} for their sample of SMGs, which are less luminous and are likely at lower redshift than our ultrared DSFGs. Therefore, the combination of previous results (both from observations and simulations) and our results suggests that a population of elliptical galaxies has been formed in intense high-redshift starbursts, represented by different SMG phases. The most massive galaxies in the local Universe formed in an SMG phase at $z\sim 4$--6, compatible with our ultrared DSFGs, whereas less massive local elliptical galaxies also formed in an SMG phase but less intense and at lower redshifts.

The measured sizes of our ultrared DSFGs are also in agreement with a scenario where these sources will evolve into the most massive ellipticals at $z \sim 3$.  Most ultrared DSFGs have physical sizes below $2.5 \, {\rm kpc}$ (there is only one source with an average size larger than $3 \, {\rm kpc}$).  The evolution from ultrared DSFGs to massive ellipticals is further supported by the analysis of the stellar populations of massive quiescent galaxies at $z \sim 2$ carried out in \cite{Krogager2014ApJ...797...17K}, which showed that the formation epoch of some of those sources is compatible with $z > 4$, comparable to the redshift distribution of our ultrared DSFGs.  The sizes of the massive ellipticals in \cite{Krogager2014ApJ...797...17K} are as large as $4.5 \, {\rm kpc}$, larger than all our ultrared DSFGs and, therefore, even the largest galaxy in our sample is compatible with a size evolution in which quiescent galaxies at $z \sim 3$ are the descendants of smaller DSFGs at higher redshifts that increase their size while or after their star formation deceases.  This evolutionary picture is also compatible with the one proposed by \cite{Barro2016ApJ...827L..32B} for the formation of compact quiescent galaxies at $z \sim 2$, which would be a scaled-down population in terms of redshift and mass of both the progenitors and descendants and luminosities of the progenitors.

\section{Conclusions}
\label{concluuuuu}

In this paper we have presented high-spatial-resolution ($\sim 0.12''$ or $\approx 800\,{\rm pc}$) ALMA 870\,$\mu$m dust continuum observations of a sample of 44 ultrared DSFGs.  These were taken from the {\it H}-ATLAS and {\it Her}MES surveys, selected to have red SPIRE colors, consistent with them being at $z\sim 4$--6. Our main conclusions are:

\begin{enumerate}

	\item We have confirmed that there exists a significant population of unlensed ultrared DSFGs which are among the most luminous sources found so far in the early Universe, forming stars at tremendous rates, up to a collective ${\rm SFR} \sim 4,500 \, M_\odot \, {\rm yr}^{-1}$ when we coadd the different sub-components into which ultrared DSFGs are resolved, and up to ${\rm SFR} \sim 2,400 \, M_\odot \, {\rm yr}^{-1}$ for individual components.

	\item The lower limits on the number counts of ultrared DSFGs at 500\,$\mu$m (where flux densities have been measured from SPIRE) conflict with models expectations, but not at 870\,$\mu$m (where flux densities have been measured from ALMA, after refinement of the samples with SCUBA-2 and LABOCA).  This can be explained by the lack of correction for flux boosting in the SPIRE flux densities, by the effect of clustering in the large SPIRE beams, or by a combination of both.  These problems do not affect ALMA, although the relatively small ALMA field of view means we might miss some emission, especially in sources with extended LABOCA emission.
	
	\item We find a variety of dust continuum morphologies, from relatively smooth disks with extended star formation to compact sources, both isolated, and interacting. The average {\sc fwhm} size of the dust continuum in our ultrared DSFGs is $\sim 1.46 \pm 0.41 \, {\rm kpc}$, so smaller than the values reported in DSFGs at lower redshifts.  
	
	\item The fact that the average size of our ultrared DSFGs is lower than that found for massive quiescent galaxies at $z \sim 2$--3 supports the idea that the former are the progenitors of the latter.  This is further supported by the expected short gas-depletion time of our ultrared DSFGs, as their star formation would cease after a few hundred million years. We are thus witnessing the birth of the high-mass end of the red sequence of galaxies.

\end{enumerate}

\begin{acknowledgements}
IO, RJI, LD and AJRL acknowledge support from the European Research Council (ERC) in the form of the Advanced Investigator Programme, 321302, {\sc cosmicism}. MN acknowledges financial support from the European Union's Horizon 2020 research and innovation programme under the Marie Sk{\l}odowska-Curie grant agreement No.~707601. IRS acknowledges support from STFC (ST/P000541/1), the ERC Advanced Investigator programme, 321334, {\sc dustygal} and a Royal Society Wolfson Merit Award. HD acknowledges financial support from the Spanish Ministry of Economy and Competitiveness (MINECO) under the 2014 Ram\'{o}n y Cajal programme MINECO RYC-2014-15686.  MJM acknowledges the support of the National Science Centre, Poland, through the POLONEZ grant 2015/19/P/ST9/04010. This project has received funding from the European Union's Horizon 2020 research and innovation programme under the Marie Sk{\l}odowska-Curie grant agreement No.~665778. DR acknowledges support from the National Science Foundation under grant number AST-1614213.  This paper makes use of the following ALMA data: ADS/JAO.ALMA\#2013.1.00001.S and ADS/JAO.ALMA\#2016.1.00139.S. ALMA is a partnership of ESO (representing its member states), NSF (USA) and NINS (Japan), together with NRC (Canada) and NSC and ASIAA (Taiwan) and KASI (Republic of Korea), in cooperation with the Republic of Chile. The Joint ALMA Observatory is operated by ESO, AUI/NRAO and NAOJ.  {\it H}-ATLAS is a project with {\it Herschel}, which is an ESA space observatory with science instruments provided by European-led Principal Investigator consortia and with important participation from NASA. The {\it H}-ATLAS website is http://www.h-atlas.org/.
\end{acknowledgements}

\bibliographystyle{mn2e}

\bibliography{ioteo_biblio}

\newpage

\appendix

\begin{figure*}
\centering

\includegraphics[width=0.90\textwidth]{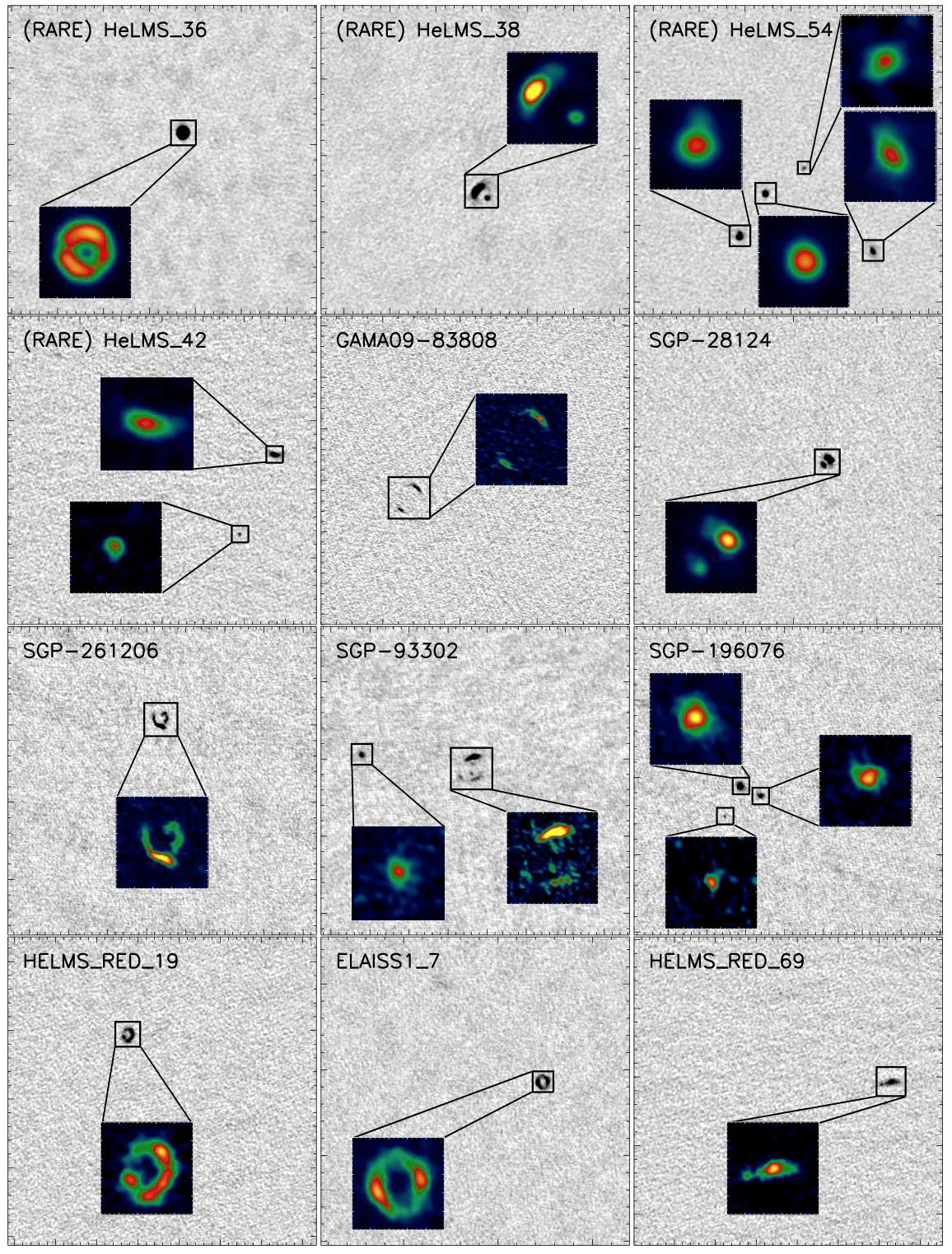} 

\caption{ALMA $870 \, {\rm \mu m}$ imaging of our UR galaxies at $\sim 0.1''$ (or $800 \, {\rm pc}$) spatial resolution. We show on each image the area covered by the primary beam to show all components which out UR starbursts are split into. On each image we include an inset figure to highlight the dust emission morphology of each detected component.
              }
\label{morphology_WIDE_map_1}
\end{figure*}

\begin{figure*}
\centering

\includegraphics[width=0.90\textwidth]{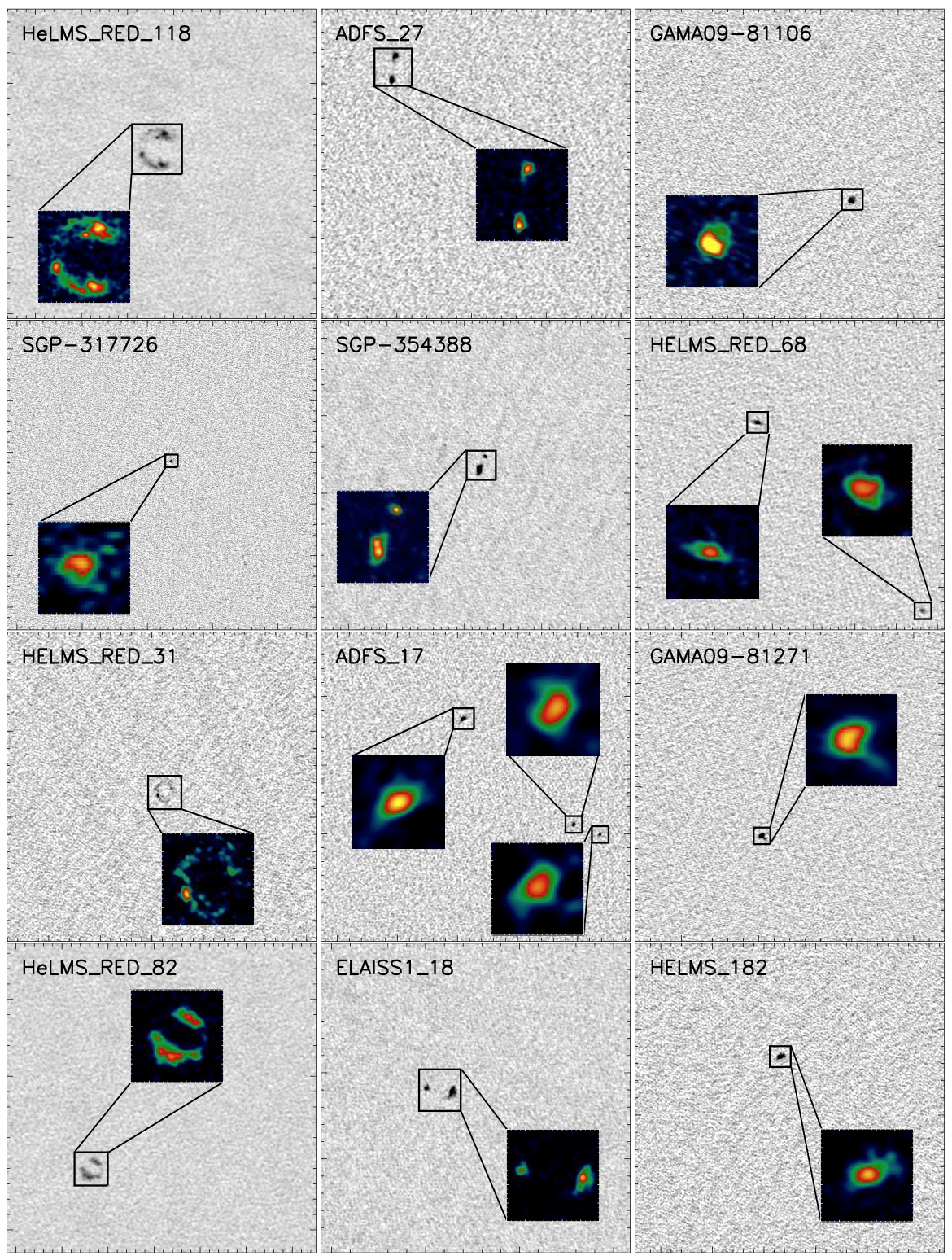}

\caption{ALMA $870 \, {\rm \mu m}$ imaging of our UR galaxies at $\sim 0.1''$ (or $800 \, {\rm pc}$) spatial resolution. We show on each image the area covered by the primary beam to show all components which out UR starbursts are split into. On each image we include an inset figure to highlight the dust emission morphology of each detected component.
              }
\label{morphology_WIDE_map_2}
\end{figure*}

\begin{figure*}
\centering

\includegraphics[width=0.90\textwidth]{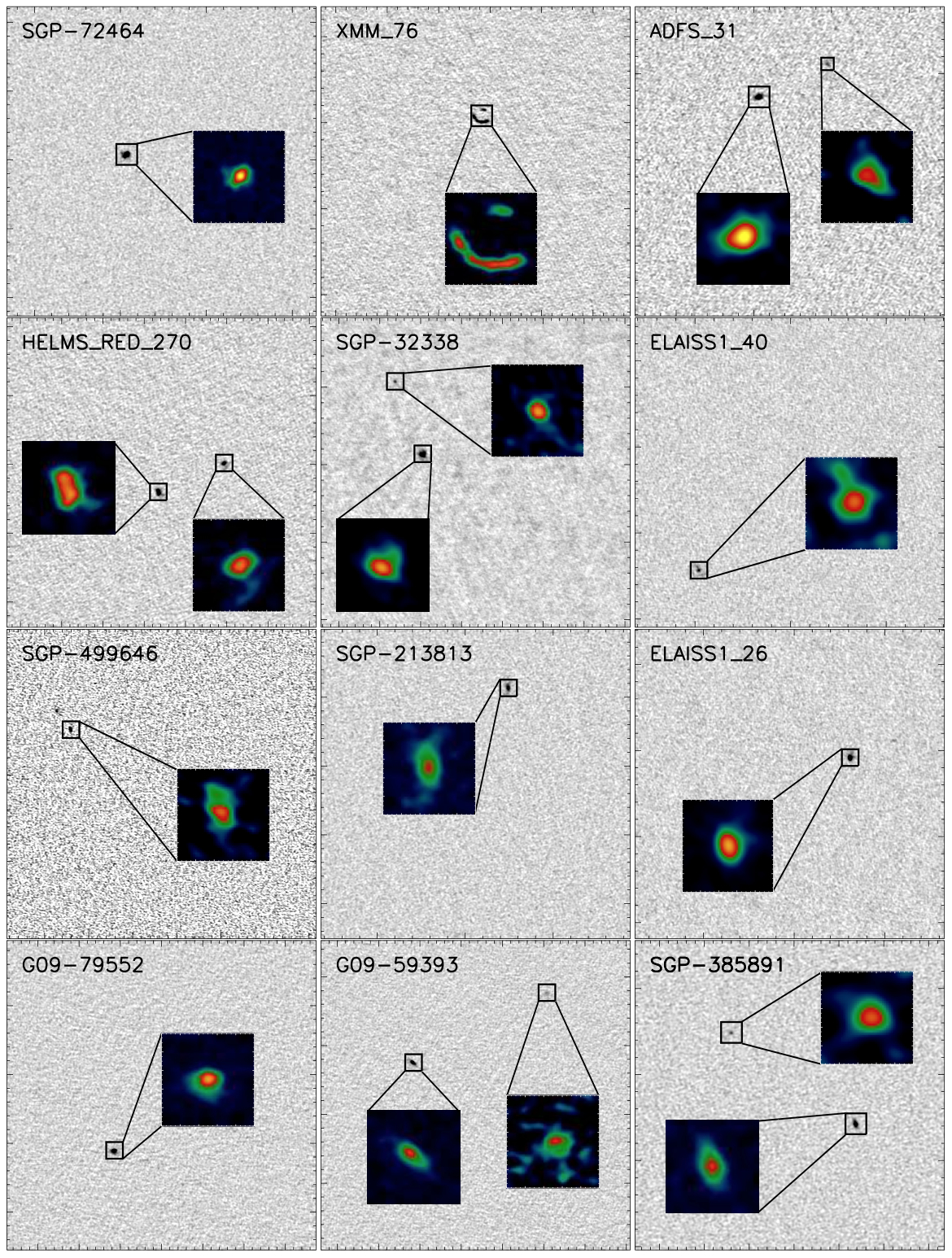}

\caption{ALMA $870 \, {\rm \mu m}$ imaging of our UR galaxies at $\sim 0.1''$ (or $800 \, {\rm pc}$) spatial resolution. We show on each image the area covered by the primary beam to show all components which out UR starbursts are split into. On each image we include an inset figure to highlight the dust emission morphology of each detected component.
              }
\label{morphology_WIDE_map_3}
\end{figure*}

\begin{figure*}
\centering

\includegraphics[width=0.90\textwidth]{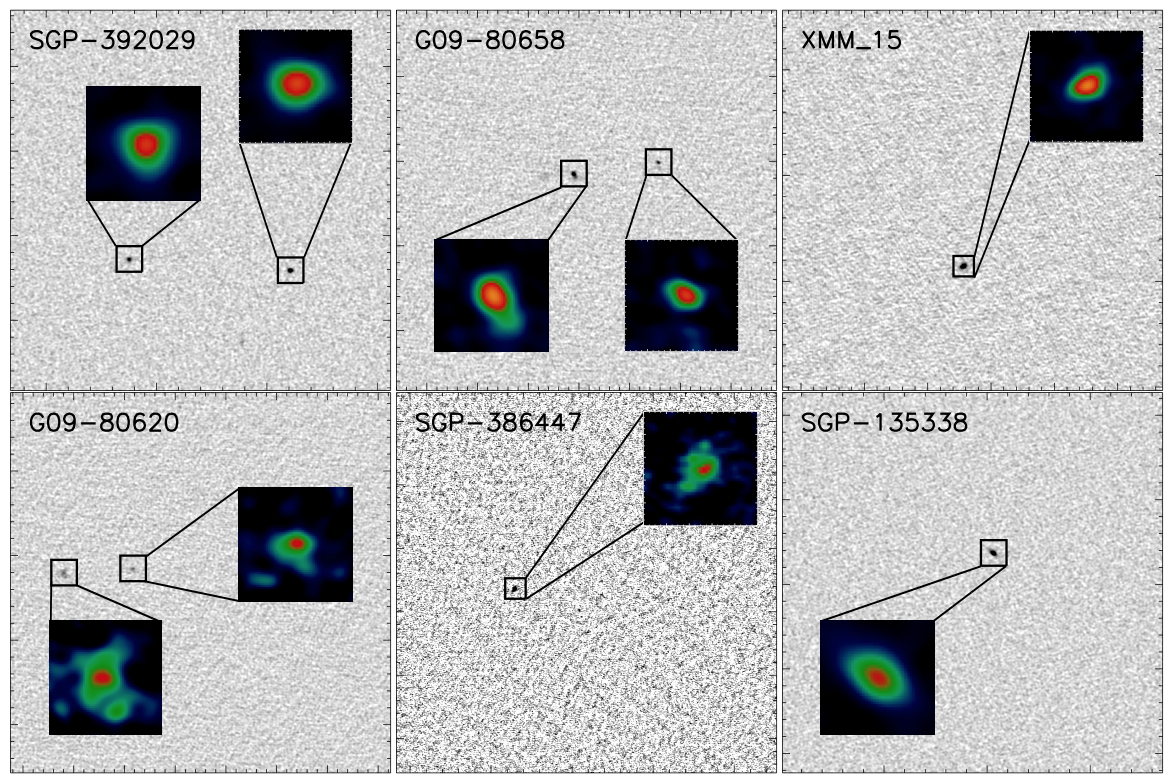}

\caption{ALMA $870 \, {\rm \mu m}$ imaging of our UR galaxies at $\sim 0.1''$ (or $800 \, {\rm pc}$) spatial resolution. We show on each image the area covered by the primary beam to show all components which out UR starbursts are split into. On each image we include an inset figure to highlight the dust emission morphology of each detected component. 
              }
\label{morphology_WIDE_map_4}
\end{figure*}

\end{document}